\documentclass[letter,12pt]{article}
\usepackage{graphicx,amssymb,amsmath,epstopdf,color,hyperref}
\usepackage{enumitem}
\usepackage{booktabs}
\usepackage{tabularx}
\usepackage{array}
\usepackage{ragged2e}
\usepackage{caption}
\usepackage[T1]{fontenc}
\usepackage[utf8]{inputenc}
\usepackage{lmodern}
\newcolumntype{Y}{>{\RaggedRight\arraybackslash}X}
\newcolumntype{C}[1]{>{\Centering\arraybackslash}p{#1}}

\def\beq{\begin{equation}}
\def\eeq{\end{equation}}
\def\bea{\begin{eqnarray}}
\def\eea{\end{eqnarray}}

\newcommand{\gsim}{\lower.7ex\hbox{$\;\stackrel{\textstyle>}{\sim}\;$}}
\newcommand{\lsim}{\lower.7ex\hbox{$\;\stackrel{\textstyle<}{\sim}\;$}}

\def\xsubsection#1{\noindent \begin{center}{\it #1}\end{center}}
\def\xxsubsection#1{\noindent {\footnotesize #1}}

\begin{document}

\noindent
\begin{center}
{\bf\large Polydoxon Transformations and Scientific Reward in Physics}

\vspace{0.3cm}
James D. Wells\footnote{jwells@umich.edu; ORCID 0000-0002-8943-5718}  \\ {\it Leinweber Institute for Theoretical Physics \\ Physics Department, University of Michigan, Ann Arbor} \\
\end{center}

\noindent
{\it Abstract:} 

We develop a descriptive account of scientific reward in physics based on the concept of the time-dependent Polydoxon, defined as the structured set of empirically viable theories at a given time. We argue that highly rewarded contributions, such as those recognized by major prizes and professional honors, can be systematically understood as those that transform this space. These transformations take the form of expansion (adding viable theories), contraction (eliminating viable theories), reconfiguration (illuminating deeper structures and relations within and between theories), and enabling moves (methodological or technological advances that enable future transformations). The analysis is further refined by emphasizing that reward correlates with the transformation's magnitude, assessed along dimensions of scope, centrality, depth, and future leverage. This framework reframes the analysis of rewarded achievement away from isolated theoretical successes and toward the dynamics of a landscape of viable theories, providing a more unified descriptive interpretation of rewarded scientific activity in physics across its diverse set of theoretical and experimental discoveries.

\tableofcontents

\vfill\eject
%%%%%%%%%%%%%%%%%%%%%%%%%%%%%%%%%%%%
%%%%%%%%%%%%%%%%%%%%%%%%%%%%%%%%%%%%%%%%%%%
\section{Introduction}

Philosophers and historians of science have long sought to characterize valued scientific work and to identify the patterns of practice that produce it. A wide range of descriptive accounts has emerged, including the influential frameworks of Popper~\cite{Popper:2002a,Popper:2002b}, Kuhn~\cite{Kuhn:1962}, Lakatos~\cite{Lakatos:1978,Lakatos:1978b}, and Laudan~\cite{Laudan:1978}, along with many recent criticisms and refinements, including~\cite{Maxwell:2012,Branahl:2025,Horvath:2023,Sepetyl:2025}. A common further step is to convert such descriptive insights into normative guidance for scientific methodology. Here, we follow the first step of this general strategy and develop a descriptive account of the type of work that is rewarded by the physics community.

The starting point of our analysis is the recognition that, in most domains of fundamental physics, there exists not a single viable theory but a vast space of theories consistent with current empirical constraints. Traditional accounts of scientific activity, which focus on the succession or competition of individual theories, do not fully incorporate this important feature from the outset. Much of the rewarded activity is better understood as operating on a large structured space of empirically viable theories, by expanding it, constraining it, or reconfiguration/reconceptualizing it through illuminating the theories within and understanding relationships between the theories within. 

Our central descriptive claim is that highly rewarded contributions, such as those recognized by major prizes and professional honors, can largely be understood as those that transform the Polydoxon\footnote{The term ``Polydoxon'' is a neologism constructed from the Greek \textit{pol\'ys} ($\pi o \lambda \acute{u} \varsigma$), meaning ``many,'' and \textit{d\'oxa} ($\delta \acute{o} \xi \alpha$), meaning ``belief.'' The suffix ``-on'' denotes a structured object. Thus, ``Polydoxon'' is intended to signify a space or collection of multiple coexisting, empirically viable beliefs or theories.}, defined as the structured set of empirically viable theories at a given time. However, not all such transformations are of equal significance. What ultimately distinguishes highly rewarded contributions is not only the type of transformation, but its magnitude, scope, and strategic importance. This refinement will be developed in detail in a later section. The support for this claim will be primarily historical: we will examine prominent examples of recognized achievement and show that they correspond to identifiable transformations of the space of viable theories. In this sense, the Polydoxon provides a unifying framework for interpreting what the community takes to be significant scientific contributions.

The Polydoxon is a theory-space, yet the set of highly rewarded contributions includes both theoretical and experimental work. This is not a contradiction. Experimental results play a central role in transforming the Polydoxon, particularly through the contraction of theory space by ruling out previously viable theories and the expansion of theory space by spurring theoretical explanations of surprising new discoveries. Experiment also participates in the reweighting of credences assigned to theories within the Polydoxon. While theoretical work is required to formulate and relate elements of the Polydoxon, experimental work is indispensable in shaping and transforming its structure. The present account is therefore theory-centric in its ontology, but it is equally experimental-centric in its valuation of scientific practice.

The novelty of the present approach does not lie in the observation that multiple theories may be empirically viable, nor in the familiar point that experiment, theory, and methodology all contribute to scientific development. Rather, it lies in the claim that these diverse forms of activity can be understood within a single structural framework as transformations of a common object, the Polydoxon of empirically viable theories, and that this perspective reveals a unifying pattern across forms of scientific achievement that are typically treated as distinct. Contributions as different as experimental discoveries, theoretical constructions, and methodological innovations can, from this perspective, be interpreted as instances of a small number of recurrent transformation types acting on theory space. At the same time, as will be emphasized later, these transformations admit a wide range of magnitudes, and it is the largest and most consequential among them that are most closely associated with high recognition. Therefore, what appears at the surface level as a heterogeneous collection of scientific advances exhibits a comparatively simple and structured form when viewed at the level of transformations of the Polydoxon.

The remainder of the paper is organized as follows. We begin by defining the Polydoxon more precisely and analyzing its structural features. We then discuss the types of transformations on the Polydoxon that scientific research brings about. We then apply historical analysis to top awards within the Physics community to show that they are generally reserved for researchers who have played a significant role in transforming the Polydoxon. Many of the detailed examples come from high energy physics and cosmology, subfields which the author is most intimately acquainted. However, all subfields fields are considered in the development of this account, and are present in the discussion categorizing Nobel Prize awards. We then turn to a more refined analysis of the magnitude of these transformations and its role in explaining differential levels of reward. We conclude with a discussion of limitations and directions for further refinement of this framework, as well as inchoate thoughts on a normative theory of physics research based on this account.

%%%%%%%%%%%%%%%%%%%%%%%%%%%%%%%%%%%%%%%%%%%%%%%%%%%%%%%
%%%%%%%%%%%%%%%%%%%%%%%%%%%%%%%%%%%%%%%%%%%%%%%%%%%%%%%
\section{The Polydoxon}

The Polydoxon is the set of all theories that are empirically viable. By theories we mean an algorithm that takes inputs and produces outputs. The inputs can be non-contingent  theory parameters, such as Yukawa couplings, gauge couplings, Planck's constant, particle content, etc., as well as contingent dynamical variables such as momentum or positions of particles, etc.  The outputs are observables, such as cross-sections, masses, decay rates, oscillation frequencies, etc. By a theory being empirically viable, we mean that there is at least one point in the input parameter space that outputs observables that are consistent with all known experiments and observations within the stated domain of the theory's applicability. By domain of a theory, we mean the space of contingent inputs that the theory will tolerate.

The above is a brief description of the Polydoxon. To make it more precise let us first discuss more what we mean by a theory, as we use the term here. We will then present a few features of the Polydoxon and its stipulated structure. And then we will give some examples of what is and is not within the Polydoxon. And to round out the discussion we will connect the Polydoxon to some other concepts in the literature, such as theory credence evaluations and pursuit worthiness.

\xsubsection{Formal expression of empirically viable theory}

More formally, we can express a theory ($T$) as a map of non-contingent static inputs (theory parameters $g$) and contingent inputs (dynamical variables $p$) within their applicability domain ($\Delta_p$) to output observables ($\sigma$): 
\begin{equation}
T:~ (g,p)\longrightarrow \sigma, ~{\rm where}~ p\in \Delta_{p}.
\end{equation}
An empirically viable theory is one in which there exists at least one choice of theory parameters $g$ that produce output observables $\sigma$ fully consistent with all known observations and experiments for any choice of $p\in \Delta_p$.

A useful point of contact for our definition of a theory is the semantic view~\cite{vanFraassen:1980,Suppe:1989,Balzer:1987,Frigg:2006,Frigg:SEP:2006,Morgan:2010}, where a theory is identified with a class of models and empirical adequacy is defined by the existence of at least one model that correctly represents observable phenomena. Our formulation of a theory is fully compatible with that perspective, but recasts it in a form that is more recognizable to practicing physicists. Rather than taking a theory to be given as a set of models, we represent it as a parameterized mapping from structured inputs, comprising both fixed theory parameters and contingent dynamical variables within a specified domain, to observable outputs. Individual models, in the language of the semantic view, correspond to particular instantiations dictated by the parameters of the theory, so that the requirement of empirical adequacy becomes the existence of at least one parameter choice yielding agreement with all observations. This reformulation does not attempt to deviate from the semantic view, but makes explicit the generative structure, parameter space, and domain restrictions that are typically implicit in physical theories. It is, in other words, an operational variant of the semantic conception rather than a departure from it.

We will generally prefer the term ``empirically viable'' in our discussion. One reason for this is because some theories compute some observables only approximately. Or, there are currently computational barriers to determining some outputs to a precision better than experimental data. Examples within the Standard Model include collider jet data and some low-energy meson and baryon decays and cross-sections. Within early universe cosmology theories, computations of some dark matter candidate's impacts on structure formation are similarly challenging.

\xsubsection{The Polydoxon is immeasurably large}

Let us consider the contributions to the Polydoxon within elementary particle physics. In this broad area of physics there are many empirically viable theories.  For example, we can add an infinity of  varieties of non-renormalizable operators to the Standard Model suppressed by the scale $\Lambda$, where $\Lambda$ is well above the weak scale. This can be classified as an infinite number of new theories, or a single new theory with an infinite number of parameters, which are the coefficients of all the gauge-invariant and Lorentz-invariant higher dimensional operators. There are also supersymmetric theories, with new superpartners at higher energies. There are theories with vector-like leptons at scales above the weak scale. There are theories of axions. Theories of dark photons. Theories of singlet Higgs bosons just around the corner from the weak scale. The ideas add up, and they are numerous. All of them have parameter space regions fully consistent with all known experimental data. 

It is no easier in cosmology. Theories of dark matter are plentiful~\cite{Yu:2025rez}. Each fitting the data within uncertainties. Theories of inflation are numerous~\cite{Baumann:2009ds}, each remaining empirically adequate after several rounds of careful observational data have been completed. More observations and experiments are planned with hopes of discerning between them, or at least ruling out some. 

%\vspace{0.5in}

\xsubsection{Structure of the Polydoxon}

From the discussion above, we realize that the first thing to know about the Polydoxon  of empirically viable theories is that it is staggeringly large. The underdetermination problem is an understatement in fundamental physics. Yet, in this infinite-sized Polydoxon, which has always been infinite, changes occur, where theories are added and subtracted. However, the Polydoxon is not just a set of theories  stacked on a pallet. There is deep structure within the Polydoxon, and understanding that structure is the source of great insight and often the precursor to future expansion or radical contraction of the Polydoxon. Let us just mention two ways that structure is understood within the Polydoxon.

First, renormalization group (RG) evolution connects theories. Multiple theories in the UV may converge to a single theory in the IR~\cite{Zinn-Justin:2021}. Theories of one type flowing into theories of another type through RG flow is a modern understanding of the structure of quantum field theories within the Polydoxon. Similarly, phase transitions that connect UV theories of one type to IR theories of another enhance the structural understanding of the Polydoxon. Another structural determination is through dualities. Two theories may look quite different to each other and otherwise would be considered independent theories within the Polydoxon, but the recognition that one is dual to the other creates insight within the theory space. For example, Randall-Sundrum warped extra dimensional theories are dual in an important sense to walking technicolor theories~\cite{Arkani-Hamed:2000ijo}, both of which were independently conceived to help solve the hierarchy problem. As another example, which we will discuss more  later, the different theories of string theory (type IIA, heterotic, etc.) and 11-dimensional supergravity are all connected by duality transformations and can be subsumed into a single theory M-theory~\cite{Witten:1995ex}. This remarkable insight reconfigures the Polydoxon in that there are edges (dualities) that connect these vertices (theories) in a graph theory sense within the Polydoxon. 

\xsubsection{Example Polydoxon theory: the Standard Model}

The Standard Model of elementary particle physics is an example of a theory within the Polydoxon. The theory parameters  include the Yukawa couplings of all the fermions, the gauge couplings of the three forces, the Higgs boson self coupling, the mass-squared of the Higgs boson, etc.~\cite{Wells:2019rwq}. The Standard Model is determined by its renormalizable lagrangian, and all the couplings of that lagrangian at some fixed scale. When computing observables the Standard Model theory may or may not need dynamical variable inputs. When computing the decay rate of the $W$ boson there are no required dynamical variables once the lagrangian is specified. However, if we wish to compute the cross-section of $e^+e^-\to W^+W^-$  scattering, for example, we need to know the momenta of the initial state electron and positron. 

Often neglected when specifying a theory is the domain of applicability of the theory~\cite{Giere:1990}. A theory is not a theory without that specification. The Standard Model is often expressed as an effective theory that is valid up to some scale $\Lambda$~\cite{Isidori:2023pyp}. In this case, no dynamical momentum or energy is allowed to be above $\Lambda$. For example, the theory is not to be trusted or used for momenta of the electron and positron above $\Lambda$.

\xsubsection{Credences and pursuitworthiness across the Polydoxon}

Within the Polydoxon some theories are much more valued than others. For example, in the Bayesian epistemological sense, the posterior of supersymmetry has gone down after the Large Hadron Collider failed to find supersymmetric partners in its data~\cite{Dawid:2024hqa}. In this sense, the credence in the theory is reduced. This leads to a lower prior on the theory for any future work. Thus we may say that theories possess priors or credences in the minds of researchers. Credences on theories can be judged by each researcher individually or by the community of scholars. There are many individual researchers whose credence for supersymmetry is as high now as it was before the LHC began taking data, but it is fair to say that the credences given to supersymmetry within the community are on average lower now. Many theories are affected by experimental results like this. 

The Polydoxon is a descriptive static object at any given moment. It is the structured collection of all empirically viable theories as described above. Strictly speaking credences are not part of the Polydoxon, as they are more subjective and can be radically different from one person to the next.  Rather it is best to consider credences as one aspect of evaluation across the Polydoxon after it has been formed, not as contributing to criteria with a bar of minimum credence/interest to gain entry to the Polydoxon.

%\xsubsection{Theories and pursuit worthy research activities}

Credence scores across the landscape are important because they help set priorities for future experimental and theoretical work. Credences are a significant input to the pursuit worthiness of activities that are centered on a particular theory or class of theories. Of course, it is only one part of pursuit worthiness, as it is well known that many other factors come into play when an individual or a community decides to pursue work centered on a particular theoretical perspective~\cite{Dawid:2013}. Other factors include how easy it is for the researcher to carry out research that aims to confirm special features of a theory or rule it out. They include perhaps the desire of a researcher or community to pursue a new direction of lower credence in order to not compete with a larger well-established community which is focused on the very highest credence theories, thereby maximizing their chances of reward.  Like credences, a sufficiently high pursuit worthiness score of activities that target confirming special features or ruling out a theory is not a criterion for a theory's admittance to the Polydoxon.  Only empirical viability  counts.

\xsubsection{Historically important and pedagogical theories}

The Polydoxon is not a Panth\'eon of famous theories. It is not a Hall of Fame for influential theories throughout history. Nor is it  a place to recognize the pedagogical value of some toy models and early simple theories. The Polydoxon is restricted entirely to be the structured set of empirically viable theories. 

For this reason, Bohr's early quantum mechanical model of the hydrogen atom is not in the Polydoxon. Nor is the plum pudding model of the atom. Nor is the Ising model. Not even Newton's theory of gravity is in the Polydoxon under the strictest definition. The latter two are interesting examples of theories that are not only famous and pedagogically rich but also useful for calculation purposes if accuracy requirements are not extreme. For example, computing trajectories of rockets using Newtonian gravity is perfectly reasonable and useful approach. The Ising model is accurate enough model in many circumstances, and represents a large coarse-grained limit of many condensed matter systems. 

There can be debate on whether or not to allow theories such as Newton's theory and the Ising model within the Polydoxon of empirically viable physics theories by specifying relaxed accuracy requirements.  In some sense, all theories are likely to be approximations or idealizations~\cite{Weisberg:2007}, and they are sure to be found wanting if experiment improves enough over time. That is a subtlety that is not fruitful for us to pursue in depth here. We merely state here that whatever theories are present in the Polydoxon they must be empirically viable under some measure of desired accuracy. Theories firmly ensconced in the Polydoxon are those that deliver output observables that conform precisely to experimental measurements within the experimental errors. Other theories, such as Newton's theory of gravity, can be members under a more relaxed accuracy requirement that is nevertheless tight and useful for scientists.

One note, however, that in practice, if we wish to be maximally strict on empirical accuracy, it is often relatively easy to determine admittance in the Polydoxon when investigating theories case by case. For example, all recognize that Einstein's theory of general relativity can be used in all circumstances that Newton's theory is used, and is always more accurate. Thus, General Relativity is securely inside the Polydoxon.

\xsubsection{The complete Polydoxon}

Finally, let us discuss the evident fact that no one individual researcher knows the full Polydoxon. Complete knowledge within a single subfield much less complete knowledge across all subfields of physics is not achievable by an individual. However, one can create charts of the Polydoxon based on individual and group knowledge within subfields.

The charts of subfield knowledge can be pieced together to form the full Poydoxon of empirically viable theories across physics. It is of little consequence that no individual can perceive the full structure for it to be a well-defined object with clear borders, just as it is of little consequence that no single individual knows the complete topography and borders of Russia to understand that Russia exists and has some notions of its extent.

Furthermore, evaluation of a theory's membership in the Polydoxon is rather objective, and scientists can easily come to agreement. This is in contrast to many other structured objects we might think are interesting, such as the ``Naturalness Set'' of all non-finetuned theories. Or, the set of ``Best Theories'', which might satisfy somebody's IBE views~\cite{Lipton:2004,Douven:2022}, for example. Deciding membership in those kinds of sets are interesting, and I would even say useful philosophy-informed discussions, for reasons discussed above. On the other hand, the ``Empirically Viable Theories Set'' is relatively straightforward to assess. Putting all of these theories together into one full structured set forms the Polydoxon at any given moment. In the next section we discuss how the Polydoxon gets transformed over time.

%%%%%%%%%%%%%%%%%%%%%%%%%%%%%%%%%%%%%%%%%%%%%%%%%%%%%%%
\section{Polydoxon transformations}
% general classification of polydoxon transformations

At any given moment there is a Polydoxon of empirically viable theories as defined in the previous section. We now wish to discuss the types of Polydoxon transformations that can take place through physics research. In this section we will be general and abstract in our identifications of Polydoxon transformations, and reserve to a later section an extensive discussion of examples of Polydoxon transformations and how the community rewards those who succeed in accomplishing it.

\xsubsection{Expanding, contracting and reconfiguring}

There are three general types of Polydoxon transformations. Through research one can add or subtract theories from the Polydoxon, and one can reconfigure theories within the Polydoxon. We can give code labels for each of these as $P_+$ (add), $P_-$ (subtract), and $P_R$ (reconfigure). 

Adding a theory to the Polydoxon can happen in two main ways. One is through creative theory work. A theorist or team of theorists devises a new empirically viable theory that has not been thought of before. Usually, the inspiration for such new theories is the attempt to solve an outstanding problem within the current Polydoxon. Or theory work may extend a theory within the Polydoxon to have a larger domain of applicability.  There are a multitude of motivations and approaches a scientists might engage in, but here we merely state that if a new theory is developed, whether it be wholly new or an extension of an existing theory, that is empirically viable the Polydoxon has been transformed by being supplemented by this new theory. This type of transformation of the Polydoxon can be symbolically represented as $P_+$, where the $+$ indicates an addition to the Polydoxon. Examples of this include Gell-Mann's theory of quarks and Einstein's theory of General Relativity.

Another way of adding to the Polydoxon is through experiment. If experiment finds new phenomena that had not been anticipated, it immediately leads to theory work that adds to the Polydoxon. Despite theory work being important for the transformation of the (theory) Polydoxon, surprising experimental work is the impetus and gets most of the ``credit'' for $P_+$.   Examples of this include the discovery of X-rays and the discovery of muons.

The Polydoxon can be contracted ($P_-$) in several ways. In pure theoretical physics, the Polydoxon can be eliminated by realizing that the purported theory is mathematically inconsistent upon closer inspection. An example of this is Pais and Uhlenbeck's higher derivative theories~\cite{Pais:1950} that were later determined to have a fatal Ostrogradsky-type instability. Those that discovered the flaws effectively eliminated such theories and so successfully contributed to $P_-$ transformations of the Polydoxon.

A more common way to contract the Polydoxon is through experimental work that rules out theories within the Polydoxon. If a theory predicts a necessary phenomenon that is not found then it is ruled out. Or, regions of parameter spaces are ruled out among theories of the Polydoxon, which contracts the Poydoxon. A radical reduction happens when a unique feature is found of a Polydoxon theory at the expense of many other theories within the Polydoxon. This radically contracts the Polydoxon by eliminating these many other theories that are suddenly found wanting. Symbolically the contraction can be represented by $P_-$ for short.  Examples include the Higgs boson discovery, which eliminated all theories in the Polydoxon that were ``Higgsless''~\cite{Wells:2018nwj}, and discovery of the size of anisotropies of the cosmic microwave background (CMB) radiation~\cite{COBE:1992syq} that eliminated many theories of structure formation primarily seeded by cosmic strings.

A third way to transform the Polydoxon is through reconfiguring it. This type of transformation primarily takes place through theory work. This is where groupings of theories or relations among theories are understood through insights of theoretical work. Such relations can be understood through dualities, for example, or common origins in gauge theories, to name two. Symbolically we can represent this reconfiguring of the Polydoxon as $P_R$. Examples of this type of transformation are the reconfiguring of understanding within the Polydoxon due Nambu's foundational work on spontaneous symmetry breaking and Wilson's theory of renormalization group.

Notice that reconfiguring the Polydoxon, $P_R$, is all about gaining deep insights into the theories that reside in the Polydoxon. In other words, it is about refining our understanding of at least some of the theories that are in the set of experimental viable theories. We might have called this ``gaining understanding'' of theories within the Polydoxon. However, our goal here is to think rather concretely about the Polydoxon, as though it is an object that must physically change. Clearly $P_+$ (adding) and $P_-$ (subtracting) change the membership of the Poydoxon set, and if each theory occupies a point in some abstract space, then the shape of the Polydoxon has changed, either grown or diminished. Reconfiguring is about the links that each point (i.e., each theory) within the Polydoxon has with all the other concepts in the world, including to other theories within the Polydoxon.

A recognized duality will make a connection between two theories in the Polydoxon. Such a connection can be visualized as a link between theories that otherwise looked different. Renormalization group flow concepts enable us to define a theory at a scale $\Lambda$, and we can compare theories at different scales and recognize that in some cases more than one distinct theory at some high scale $\Lambda$ will flow to the same theory in the deep infrared $\Lambda\to 0$. This IR equivalence between theories forms a new recognized bond within the Polydoxon. Or, take for example, the early theories of quantum mechanics that introduced spin, but only later was it understood that spin was deeply connected to Einstein's theory of relativity through representations of the Lorentz group. Such deepening of understanding forms bonds between concepts within empirically viable theories that reconfigures the Polydoxon. The recognition of principles in common across the Polydoxon are valuable $P_R$ activities as well, such as the illumination of the CPT theorem, Newton's laws of motion, symmetries, etc.\ The principles are shared among many theories and helps create conceptual links within the Polydoxon.

To generalize the definition of reconfiguring the Polydoxon we consider a $P_R$ transformation to have taken place if there is  a change in understanding or perspective of empirically viable theory(ies) within the Polydoxon through some link, recognized shared principles, or newly realized relationships to concepts that were heretofore external to the theory.

\xsubsection{Theory and experimental cooperative research}

These three operations $P_+$, $P_-$ and $P_R$ are put forward as the three distinct ways that the Poydoxon of empirically viable theories can be transformed. It is important to note that in almost all of them deep cooperation is needed between theoretical physics research and experimental research. The most radical transformations generally happen when high levels of creativity and grit are present in the theoretical and experimental work behind it. Although the Polydoxon definition is a theory-centered definition, what really constitutes rewarded success are the {\it changes} of the Polydoxon, and experiment is indispensable and equally central to the endeavor. 

One should also remark that the phrases ``theoretical research'' and ``experimental research'' are used here rather than ``theorists'' and ``experimentalists.'' The reason is that good theorists who are transforming the Polydoxon do both theory and experimental work, albeit primarily theory. And good experimentalists transforming the Polydoxon engage in both experimental and theoretical work, albeit primarily experimental. Cooperation between these two sides not only must be smooth among researchers within the community, but within each of the individuals themselves.

\xsubsection{Enabling moves}

In addition to these types of work that directly transform the Polydoxon there is one that is indirect. These we call ``enabling moves.'' An enabling move is a contribution that does not yet transform the Polydoxon, but changes the epistemic, computational, or experimental conditions so that a substantial transformation becomes newly possible or imminent.

An enabling move is somewhat similar to a ``quiet move'' in chess, which denotes a move that captures no pieces and does not seem to the uninitiated to bring immediate value to the player but is nevertheless a highly valuable move that positions the player well for future captures and even victory. In business such actions are sometimes called ``strategically positioning'', where a company might even lose money in order to make much more money later. 

Examples of enabling moves include, for example, innovations in tracking (e.g., silicon vertex detector~\cite{Heijne:1980rt}) at a particle collider and the development of non-Abelian Yang-Mills theories~\cite{Yang:1954ek}. One of the most consequential enabling moves in recent years was the discovery of gravity waves~\cite{LIGOScientific:2016aoc} which we will discuss below. We will symbolically label these enabling moves as $P_E$.

At first glance, $P_R$ (reconfiguring) and $P_E$ (enabling moves) may look similar. However, there is a key difference. $P_R$ is about restructuring our formulation of theories and their connections within the Polydoxon. In other words, it is about gaining deeper understanding of theories and relationships between theories within the Polydoxon. $P_E$ does not change theories within the Polydoxon or our understanding of them, but rather uses expert knowledge of the Polydoxon  to set us up to make future transformations. It is an expertly placed indirect kick or assist that leads to a goal, to use a soccer analogy. It is a new technology, a new analysis method, or a new mathematical innovation, for example, which sets up the community to directly transform the Polydoxon through $P_+$, $P_-$, or $P_R$.

An example of a powerful enabling move ($P_E$) in physics is Bell's development of Bell's Inequalities. John Bell showed~\cite{Bell:1964kc} that there is an inequality of spin projection measurements that is necessary in any local hidden-variable theory of quantum mechanics, but which is violated in standard quantum mechanics without hidden variables. To a lesser extent Bell's work restructured the Polydoxon ($P_R$) as well, in that it enabled one to divide theories into two classes: those that violate and those that do not violate Bell's inequalities. More importantly, however, this enabled decisive experimental tests of such hidden variable theories, culminating in the experimental work of Clauser~\cite{Clauser:1972}, Aspect~\cite{Aspect:1982}, and Zeilinger~\cite{Zeilinger:1998}, who showed violations of Bell's inequalities, thereby expelling ($P_-$) a large class of local hidden-variable theories from further consideration. Enabling moves such as these often have the effect of turning philosophical disputes into experimental programs. For this substantial enabling move John Bell was awarded the Dirac Medal in 1988, and for their experimental work on contracting the Polydoxon Clauser, Aspect and Zeilinger won the Nobel Prize in physics in 2022.

In summary, the direct and indirect transformations on the Polydoxon are $P_+$, $P_-$, $P_R$, and $P_E$, which represent, respectively, adding to, subtracting from, and reconfiguring the Polydoxon, and performing enabling moves.

\xsubsection{Connections to Kuhn}

Many physicists are well-versed in the language of Thomas Kuhn's {\it Structure of Scientific Revolutions}~\cite{Kuhn:1962}. For this reason we give a few brief comments that compare and contrast the account given here with that of Kuhn. For Kuhn, a subfield of science is dominated by a leading paradigm, wherein scientists do a significant amount of research that prods and probes the theory in many directions, and uses the theory for interpretation of phenomena. When conceptual puzzles and anomalies increase, some scientists break from the prevailing paradigm and propose a new paradigm that solves the new puzzles while retaining past successes. This is the process of revolution toward a new paradigm.

The framework presented here, on the other hand, is centered on the landscape of empirically viable theories rather than a dominant paradigm. For us, normal science work shows up in the categories of reconfiguring the Polydoxon ($P_R$) and enabling moves ($P_E$) which are most often carried out within the confines of the current Polydoxon but positions the field to reach beyond it. Revolutionary physics is characterized by novel expansions ($P_+$) and radical contractions ($P_-$) of the Polydoxon. These expansions do not necessarily replace the dominant theory. In each of these categories there are gradations on the level of impact. For Kuhn, both normal science and revolutionary science can be incremental or exceptional, just as for us these transformations can be incremental ($\delta P_i$) or exceptional, as we will discuss in a later section.

%%%%%%%%%%%%%%%%%%%%
\section{Rewards for Polydoxon transformations}
% examples of scientific awards are for mostly for polydoxon transformations 

Up to this point we have defined the Polydoxon of empirically viable theories, and we have described various types of transformations that can take place with respect to the Polydoxon. We now wish to make the connection between the Polydoxon and physics awards. Our central descriptive claim  is that rewarded research activity in physics is generally due to achieving substantial Polydoxon transformations ($P_+$, $P_-$, $P_R$  and $P_E$).  We wish to support this claim with examples in this section. As we will see, the type of activity that is highly rewarded within physics that is not centered on transforming the Polydoxon is society-impacting engineering springing out of the physics tradition.

\xsubsection{Rewards for what?}

Our descriptive account attempts to give greater organizational structure and clarity to the type of research that the physics community gives its top research awards. However, it must be acknowledged that there are many types of awards that are given out to physicists. Some awards are for outstanding research, and other awards are given for non-research activities, e.g., helping the public to understand better the goals and achievements of science, or helping policymakers understand the need for and the implications of scientific work.

For the top awards that target high-impact research, it is worth noting that they do not specify precisely what they are looking for. The descriptions for those who would qualify are broad and somewhat vague. For example, here are the critieria of several top awards within physics:

Nobel Prize: ``... to the person[s] who shall have made the most important discovery or invention within the field of physics''~\cite{Nobel:will}

Sakurai Prize: ``to recognize and encourage outstanding achievement in particle theory.''~\cite{SakuraiPrizes}. 

Panofsky Prize ``to recognize and encourage outstanding achievements in experimental particle physics.''~\cite{PanofskyPrizes}. 

The APS Medal: ``The APS Medal was established to recognize contributions of the highest level that advance our knowledge and understanding of the physical universe in all its facets.''~\cite{APS:Medal}. 

The ICTP Dirac Medal: ``It is awarded ... to scientists who have made significant contributions to theoretical physics.'' ~\cite{DiracMedal}

Breakthrough Prize: ``to recognize those individuals who have made profound contributions to human knowledge.''~\cite{BreakthroughPrize}

Member of National Academies of Science~\cite{NAS:members} : ``Members are elected to the National Academy of Sciences in recognition of their distinguished and continuing achievements in original research.''

The words and phrases used here to describe what receives one of these top awards in physics include ``important discovery'', ``outstanding achievement'', ``contributions of the highest order'', ``advance knowledge and understanding'', ``significant contributions'', ``profound contributions'', ``achievements'', etc. These are not highly prescriptive phrases. Rather, they are calls to scientists to point out and recognize what research work qualifies under these majestic phrases. Necessarily a significant amount of intuition is employed that cannot easily be articulated in a general algorithmic sense.

%\vspace{0.4in}

\xsubsection{Scrutiny and group decisions}

The focus of our work here is entirely a descriptive one: what work gets rewarded by the top prizes in physics? It is in principle a different question to ask, what work was the most important? There can be numerous responses that differ. Nevertheless, the major awards are the focus here for several reasons. First, it is something to analyze objectively through their records. Second, they are desired by researchers due to its bringing prestige, money, and satisfaction to the winners. And third, it is the good reference point toward supporting a normative theory of scientific research that aims to reveal activities and approaches that lead to maximum impact.

Regarding the third point, we will have little to say here, as we do not intend to develop the normative aspect. However, we remark only that a key finding of social epistemology is that groups of people with their reputations at stake together can make better decisions than any one individual (see, e.g.,~\cite{Marshall:2017,Tindale:2019}). There are well-known potential distortions to group decision making (see \cite{Berger:1972,Rossiter:1993} and the many followups), such as prestige dynamics, supporting allies, promotion of subfield research directions, sexism, etc., but these distortions do not substantively divert us from the conclusion that the Nobel Prize and APS Prizes are better indicators of research value than any other known constitutive group, including even the opinions of a single highly esteemed researcher. 

To emphasize further, the Nobel Prize of Physics is one of the most scrutinized awards in the world~\cite{Nobel:nomination,Chen:2023}. There is substantial care taken to have excellent physicists across multiple subfields of physics collect nominations, have discussions, and make selections that have broad support. Institutional reputation is also at stake, motivating committees to not make selections that are roundly criticized. Although criticism cannot be escaped, even legitimate criticism~\cite{Zuckerman:1977,Rossiter:1993}, it is difficult to find a process of recognition that is better. Similarly, albeit it to a lesser degree, the APS prizes have similar rigor~\cite{APS:Medal}, and constitute a community-wide judgment of valuable research more reliably than any one physicist could do. 

Let us not discuss this aspect further here, as this risks wading too deep into normative waters for our present purposes. Let us get back to our descriptive account that research activities that receive major awards are those that transform the Polydoxon, or perform enabling moves that lead to imminent transformations of the Polydoxon.
In what follows are some prominent examples of rewarded physics work, and their impact on the Polydoxon of empirically viable theories.

%\vspace{0.5in}

\xsubsection{Examples}

\xxsubsection{Precision Z-pole theory and experiment}

In the late 1980s experiments at CERN and at SLAC began operating machines that collided electrons with positrons on the $Z$ boson resonance. The $Z$ boson then decayed into various possible final states, such as $e^+e^-$, $b\bar b$, $\tau^+\tau^-$, $\nu\bar \nu$, etc. The Standard Model of particle physics had only been established as the preeminent theory of elementary particles the decade before, and these ``$Z$-pole experiments'' were commissioned to study very precisely exactly how these newly discovered heavy $Z$ bosons would decay. There were many ideas of physics beyond the Standard Model that predicted that the probabilities of the $Z$ boson decaying into its different possible final states would differ enough from the Standard Model that the CERN and SLAC experiments would discern it. 

A tremendous theoretical effort was undertaken to compute these probabilities at higher quantum loop order. The combination of this experimental work and experimental work was able to rule out many theories of physics beyond the Standard Model, in particular most so-called technicolor theories. By virtue of this theory and experimental work, the Polydoxon contracted significantly, leaving the Standard Model standing and a few other notable theories, such as low-scale supersymmetry. 

Many researchers that were critical for the success of this program received high awards from the APS. For example, Alberto Sirlin and William Marciano won the 2002 Sakurai Prize ``for their pioneering work on radiative corrections, which made precision electroweak studies a powerful method of probing the Standard Model and searching for new physics.''  On the experimental side, Martin Breidenback won the 2000 Panofsky prize for his work that led to ``important advances both in the measurement of electroweak parameters and in accelerator technology.''

\xxsubsection{Warped Extra Dimensions}

In the late 1990s there was a flurry of activity on theories of extra spatial dimensions. One variant of that is warped extra dimensions, which enables a warping factor difference between an extra dimension and our normal three dimensional space. This can dramatically reduce the otherwise inexplicable large hierarchies of mass scales, such as the Planck scale to the weak scale. It is a speculative idea but it was immediately recognized as a theory that is empirically adequate and that could solve the hierarchy problem. This is clearly a case of expanding the Polydoxon in a dramatic fashion.

The original authors of warped extra dimensions, Lisa Randall and Raman Sundrum~\cite{Randall:1999ee}, were jointly awarded the Sakurai prize of 2019 ``for creative contributions to physics beyond the Standard Model, in particular the discovery that warped extra dimensions of space can solve the hierarchy puzzle, which has had a tremendous impact on searches at the Large Hadron Collider.''~\cite{Sakurai-Randall-Sundrum}.

\xxsubsection{Neutrino mass discovery}

For many decades it was thought that the neutrino masses were zero. There was no experiment that could detect any value of its mass not consistent with zero~\cite{Bilenky:2012qb}. Then anomalies arose over time. There was a deficit of electron neutrinos arriving from the sun compared to what solar models expected. There was a deficit of muon flavored neutrinos coming from atmospheric neutrinos (those that arise from pions in cosmic rays). A review article from the early 1970s suggested that the evidences for mass may be experimental error~\cite{Trimble:1973ca}. But it was not. In time experimentalists determined that neutrinos had to have mass in order to explain all the data. 

The discovery of neutrino mass had a profound impact on the Polydoxon. All theories that necessitated neutrinos have zero mass were banished permanently from the Polydoxon. This was a radical contraction, especially since the only theories that researchers thought were reasonable ones for so many years were zero-mass neutrino theories. Furthermore, it opened up new areas of inquiry where theorists were pushed to come up with numerous different theories for how the neutrino could get mass. These new ideas expanded the Polydoxon, which remains bloated with neutrino mass theories to this day. The hope is that new experiments in the future will be able to contract the Polydoxon by eliminating many of those theories.

The discovery of neutrino masses has yielded a Nobel Prize and many APS prizes. We cite just the Nobel Prize awarded here. The 2015 Nobel Prize was awarded to Takaaki Kajita and Arthur McDonald ``for their key contributions to the experiments which demonstrated that neutrinos change identities. This metamorphosis requires that neutrinos have mass.''~\cite{NobelPrizes}.

\xxsubsection{Higgs boson discovery}

In 2012 the Higgs boson was discovered. Said more precisely, a scalar boson that is consistent with being the Standard Model's Higgs boson was discovered. This discovery was covered extensively in the press and was considered by many scientists one of the most monumental discoveries in decades. The excitement of this discovery can be interpreted as a radical contraction of the Polydoxon. 

Before its discovery, the Higgs boson was a speculative particle. There was much skepticism that such a simple scalar boson existed~\cite{Wells:2018nwj}. It was recognized that nature so far had not presented us with any fundamental scalar boson particles. The Higgs boson would be the first. For that reason, and other reasons having to do with solving the Hierarchy Problem between the Planck scale and the weak scale, there were numerous other theories developed that eschewed a single elementary Higgs boson like that of the Standard Model.

Once the Higgs boson was discovered, all theories that disallowed the Higgs boson, such as the aptly named ``Higgsless theories'', were immediately banished from the Polydoxon. This large contraction was the primary source of excitement. The secondary source of excitement was that once it was found, one could do precision experimental analysis of its decays at the Large Hadron Collider and other future colliders to confirm some areas and banish other areas of the Polydoxon. 

Numerous awards have been given for the Higgs boson discovery. The Nobel Prize to theorists Francois Englert and Peter Higgs was awarded in 2013 ``for the theoretical discovery of a mechanism that contributes to our understanding of the origin of mass of subatomic particles, and which recently was confirmed through the discovery of the predicted fundamental particle....''~\cite{SakuraiPrizes}. Experimentalists were awarded the Panofsky prize in 2017 ``for distinguished leadership in the conception, design, and construction of the ATLAS and CMS detectors, which were instrumental in the discovery of the Higgs boson.''~\cite{PanofskyPrizes}.

\xxsubsection{Gravity waves discovery}

In 2015 the LIGO collaboration detected gravity waves directly for the first time. It was a signal produced by the merger of two black holes, one 36 times the mass of our sun, and the other 29 times more massive. This discovery was hailed as one of the great discoveries of this century. It resulted in the Nobel Prize being awarded to Rainer Weiss, Barry Barish, and Kip Thorne ``for decisive contributions to the LIGO detector and the observation of gravitational waves.''~\cite{NobelPrizes}.

At first glance it does not have much to do with transformations of the Polydoxon. In fact, the existence of gravity waves was completely expected~\cite{ChapelHill:1957}. Unlike the Higgs boson, it was not doubted by anyone that if one built an experiment with enough sensitivity one would see it. Therefore, we must ask: why the excitement? And, what does this have to do with Polydoxon transformations?

The answer is what we called above an enabling move. What made the gravity wave detection so transformational to the field is that it proved that we have completed the work necessary to probe and test ideas of early universe cosmology that were unthinkably inaccessible before. Theories that were safely empirically viable are now faced with the pressure of checking their gravitational wave impacts. Theories entering a new phase of experimental risk include early universe phases transitions, copious supermassive black hole production, and  some dark matter theories to name a few. 

It is instructive to compare the discovery of gravity waves to the discovery of the tau neutrino. There was no Nobel prize for the tau neutrino. There was no worldwide celebration and cheering. There was no explosion of awards for the scientists connected to the tau neutrino discovery. Why would that be the case since the expectation of the tau neutrino existing was no different than the expectation that gravity waves existed? The difference is that gravity wave detection was immediately recognized to have dramatic value in transforming the Polydoxon of empirically viable speculative theories. It was one of the greatest enabling moves in experimental history, on par with the invention of high-energy colliders to explore the frontier of elementary particles.

\xsubsection{Nobel Prizes and Polydoxon Transformations}

So far we have discussed several high profile discoveries of theoretical physics and experimental physics, and shown how they can be interpreted as significant transformations of the Polydoxon. The compilation of these anecdotes signals that Polydoxon transformation indeed are important, but it does not prove that Polydoxon transformations are the only rewarded activities in physics. It is here where I wish to make that case by looking carefully at Nobel Prizes that have been awarded in physics, and ask if they are all for Polydoxon transformations.

Nobel Prizes in Physics have been awarded since 1901. The first Nobel Prize went to Wilhelm R\"ontgen for discovery of ``remarkable rays'', which today we call X-rays. This discovery of new phenomena classifies rather directly as adding to the Polydoxon, as it induced immediate and extensive theory production to account for the new phenomena, adding new theories to the Polydoxon, which gained further support or were rejected later by additional theory and experimental work. 

This first Nobel Prize of a serendipitous  experimental discovery is the quintessential science discovery leading immediately to Polydoxon expansion. There are other such discoveries, including radioactivity (the Curies, 1903), cosmic microwave radiation (Penzias and Wilson, 1978), Cosmic rays (Hess, 1936), Superconductivity (Onnes, 1913), the positron (Anderson, 1936), Giant Magnetoresistance (Fert, Gr\"unberg, 2007), tau lepton (Perl, 1995), accelerating universe (Perlmutter, Schmidt, Riess, 2011), and Graphene (Geim, Novoselov, 2010). In most of these cases, there was an explosion of theory addition to the Polydoxon. In the case of the positron, without Anderson immediately recognizing it,  the discovery confirmed a key aspect of Dirac's quantum theory of the electron, thus contracting the Polydoxon by expelling competitor theories.

Experimental results that confirm key features of a theory in the Polydoxon at the expense of other theories in the Polydoxon are examples of discovery leading to radical reduction of the Polydoxon by banishing those competitor theories. The positron discovery (Anderson, 1936) is one case we discussed above. Other examples with various degrees of expectation are the Higgs boson (Englert and Higgs, 2013), Compton effect (Compton, 1927), muon neutrino (Lederman, Schwartz, Steinberger, 1988), quantum hall effect (von Klitzing, 1985), deep inelastic scattering (Friedman, Kendall, Taylor, 1990), neutrino detection (Reines, 1995), Electron diffraction (Davisson, 1937), and binary pulsar spin down (Hulse, Taylor, 1993).

Several discoveries confirming key features of a theory were more or less entirely expected and there was somewhat less direct effect on the Polydoxon. One key example of this is the discovery of the   $W$ and $Z$ bosons (Rubbia, van der Meer, 1984). But as we discussed earlier in this section for gravity waves, each of these signaled a qualitatively new era of experimental power and were powerful enabling moves that accomplished strategic positioning for future transformations of the Polydoxon. The discovery of the $W$ and $Z$ heralded an era of hadron supercolliders that would bring so many powerful probes to transform the Polydoxon for decades to come.

Looking across more than a century of Nobel prizes they are very well classified as creative theory additions to the Polydoxon, surprising new experimental discoveries (leading to expansion of Polydoxon), strong experimental support for some theory(ies) over others in the Polydoxon (leading to contraction of the Polydoxon), and experiment that strategically positions the field for future Polydoxon transformations (enabling moves). 

There are many strategic positioning activities that have been rewarded with a Nobel Prize. These are focused on new techniques, experimental and computational, that hold strong promise of being used for future transformations of the Polydoxon. For example, chirped pulse amplification (Mourou, Strickland, 2018), optical tweezers (Ashkin, 2018), CCD sensor (Boyle, Smith, 2009), maser-laser (Townes, Basov, Prokhorov, 1964), quantum control (Haroche, Wineland, 2012), machine learning (Hopfield, Hinton, 2024), semiconductor heterostructures (Alferov, Kroemer, 2000), and electron microscope (1986).

%\xsubsection{Rewards for reconfiguring}

Several Nobel prizes have been awarded for deepening our understanding of the theories within the Polydoxon and thus firming up or rearranging its structure in the process. This is accomplished by drawing attention to connections between theories or understanding the behavior of the theories in different dynamical ranges, thus enabling more careful crucial tests of the theories. 
Some Nobel Prizes in this category are asymptotic freedom (Gross, Wilczek, Politzer, 2004), spontaneous symmetry breaking (Nambu, 2008), topological phases and classifications (Thouless, Haldane, Kosterlitz, 2016), renormalization group (Wilson, 1982), renormalizable gauge theories (`t Hooft, Veltman, 1999), superfluidity (Landau, 1982), Anderson localization (Anderson, 1977), black hole singularities (Penrose, 2020).

\xsubsection{Mathematical physics}

Mathematical physics is a particularly difficult category of research to judge for awards within the physics community. By mathematical physics we mean mostly the application of advanced mathematics to real or speculative physical systems. The most recognized contributions to mathematical physics are those that introduce a particular element of mathematics to bear on a physical system in a manner not previously recognized. Group theory for elementary particle physics, topology for condensed matter physics, differential geometry for general relativity and gauge theories, category theory for quantum mechanics and quantum information,  etc.

Some of these creative mathematical connections have led to revolutionary understanding of fundamental physics, such as Lie groups and Lie algebras efficiently used in formulations of fundamental physics theories. However, the connection may be made only a long time into the future, or not at all, which makes rewarding such activity particularly difficult. As we shall see below, Robert Mills was never rewarded for bringing non-Abelian groups to gauge theories because the precise theory he and Yang constructed was not empirically adequate. It was not until many years after Yang and Mills 1954 paper~\cite{Yang:1954ek} that non-Abelian gauge theories found empirically viable applications.

To counter such gaps, some awards within physics have been constructed to encourage deep mathematical physics thinking, knowing that it is a necessary subject with uncertain and perhaps even low payoff with respect to the highest level prizes of recognition, such as the Nobel Prize. The Dannie Heineman Prize is one such award, which ``recognizes outstanding publications in the field of mathematical physics.''~\cite{Dannie:APS}. For example, the 2023 recipient Nikita Nekrasov was awarded the Dannie Heineman Prize for ``the elegant application of powerful mathematical techniques to extract exact results for quantum field theories, as well as shedding light on integrable systems and non-commutative geometry.'' None of the theories worked on are empirically viable in the form studied, and therefore do not transform the Polydoxon. And because there is no sense in which the work renders inevitable significant transformations of the Polydoxon it does not qualify as an enabling move. Therefore, according to our account, then, such work, and other similar type of mathematical physics work, is not eligible for the very highest prizes, such as the Nobel Prize. Similar considerations apply to  other winners of the Dannie Heineman prize, including another recent awardee Samson Shatashvili who was awarded, in part, for ``the co-discovery of Bethe/gauge correspondence between supersymmetric vacua and quantum integrability.'' These are activities rather distant from the Polydoxon. The hope, which may be long in being realized, is that these examples and similar mathematically oriented work will turn one day into powerful techniques directly applicable to empirically relevant theories. 

\xsubsection{Rewards for engineering}

A few Nobel Prizes do not fit well within the Polydoxon transformation categories. They include transistor (Bardeen, Brattain, Shockley, 1956), integrated circuit (Kilby, 2000), optical fibers (Kao, 2009), and the blue LED (Nakamura, 2014). These are all engineering feats, demonstrating extraordinary mastery of physical systems. They each have had profound and direct societal impact to such a high degree that they were impossible to ignore. Nevertheless, they are awards for engineering and not fundamental science, as many physicists would characterize it. For example, APS news characterized the blue LED Nobel Prize of 2014 as an award for those who ``solved several major engineering problems''~\cite{APS:2014}.

\xsubsection{Summary of reward categories}

In summary, there are several categories under which we can classify all Nobel Prize awards. I will list them, give one example in parenthesis, and state how they affect the Polydoxon:
\begin{itemize}
\item Unexpected experimental discoveries (X-rays) $\longrightarrow$ Polydoxon expansion ($P_+$)
\item Theory creation (electroweak theory) $\longrightarrow$ Polydoxon expansion ($P_+$)
\item Discerning experimental discoveries (Higgs boson) $\longrightarrow$ Polydoxon contraction ($P_-$)
\item Expected but challenging experimental discoveries signifying new era (gravity waves) $\longrightarrow$ enabling moves that hold promise for Polydoxon transformations ($P_E$)
\item Experimental, theoretical or computational tools and methods (cloud chamber) $\longrightarrow$ enabling moves that hold promise for Polydoxon transformations ($P_E$)
\item Theory analysis (asymptotic freedom) $\longrightarrow$ Polydoxon reconfiguring/clarification ($P_R$)
\item Transformative engineering (blue LED) $\longrightarrow$ no effect on the Polydoxon (Engineering)
\end{itemize}
Except for the engineering category, of which there are very few Nobel Prizes in Physics, a strong case can be made for each category contributing to substantial Polydoxon transformations on the structured landscape of empirically viable theories.

\xsubsection{Theory risks and rewards}

We have shown that the highest rewards are given generally to those who have transformed the Polydoxon. An award is a judgement on results, with little consideration of the research methodological path the researcher took to arrive at that accomplishment. However, for all research programs that intend to have surprising and high payoffs, the initial steps  involve risks that the effort does not pay off, the Polydoxon is not transformed, and the researcher receives little or no reward for their efforts.

The most consequential discoveries often begin with researchers not knowing if their inchoate efforts toward some goal will pay off. The process is a creative one to generate ideas that solve problems, and then it is an analytic exercise to determine if their creative ideas actually work and are compatible with all known experiment. Sometimes the creative researcher is not an expert in the analytic part, and may not know well the panoply of relevant experimental data to the analysis that determines if the theory idea is empirically viable. In either case, the time it takes to go from creative spark to entry into the Polydoxon, if the resulting theory survives, may be long and involve multiple research programs with complementary skills.

Let us discuss two examples of this. One is the theory of large extra dimensions~\cite{Arkani-Hamed:1998jmv,Antoniadis:1998ig}. The theory idea was very clever, and it took little time to write down the theory once the creative spark happened. However, the analytic part -- trying to determine if the theory was compatible with all known experiment -- was more difficult. In particular this was because the researchers were interested in large extra dimensions solving the hierarchy problem, which put the scale of these large extra dimensions very low. Gravity could be altered at distances as large as 1 mm, which is a distance one can see with the naked eye on a ruler. It happened in that case that the researchers were also expert phenomenologists, and were able in time to do much of the analytic work themselves to establish its empirical viability and give confidence of its admittance to the Polydoxon. Neverthless, it was a risky endeavor to invest the time to establish that, when it was uncertain at the beginning. 

Another example is the theory of non-abelian gauge symmetries introduced in 1954 by Yang and Mills. These researchers were interested in solving a theoretical puzzle. It appeared that the proton and neutron were two parts of an isospin doublet. They had very similar masses and their cross-sections gave support to it. However, the idea was not well suited to relativistic local quantum field theory, and they wanted to find a purely local theory, meaning isospin rotations can be independently performed at each spacetime point. What resulted was non-abelian gauge symmetry.

We know today that non-abelian gauge symmetry is the description of the electroweak and strong interactions. In that sense the idea of introducing it by Yang and Mills was a milestone in theoretical physics. However, the theory they introduced was simply wrong. The strong interaction was known to be a short-range force whereas Yang and Mills's theory predicted long range. Furthermore, the proton and neutron are not members of doublet of a gauge symmetry representation but rather a global flavor symmetry. The specific theory presented by Yang and Mills never entered the Polydoxon. 

For this reason, even though Yang-Mills theory is taught as one of the central tenets of modern quantum field theory, Mills did not receive a major international awards for this work. Many awards came later to others for the firm establishment and understanding of non-abelian field theories as part of the Polydoxon (Gell-Mann, Weinberg, Salam, Glashow, 't Hooft, Veltman, Gross, Wilczek, Politzer to name the theorists). Mills being frozen out of such awards is to be accounted for here due to  never having expressed a Polydoxonic truth. From the point of view of rewards, Mills's effort was risky and did not pay off. Yang received many future awards, and the Nobel Prize, but not for Yang-Mills theory. Rather, his work on parity violation, which did enter the Polydoxon, gained these highest rewards. 

%\section{Productive rewarded activities}

\xsubsection{Polydoxon-neutral activities}

The identification of rewarded activities described above has the implication that  activities that have little prospect of transforming the Polydoxon go largely unrewarded.

The first Polydoxon-neutral activity we wish to point out is experimental work that purposely ignores the current Polydoxon and conducts experiments that do not put at risk any theory within the Polydoxon. Such experiments may be lucky and find something dramatically new, which would transform the Polydoxon. However, there is an infinite space of experiments one could conduct that are not ``crucial experiments", meaning they are guaranteed to have impact on the Polydoxon. There is no known method that enhances serendipitous discovery by purposely targeting a class of non-crucial experiments, which do not put any theory within the Polydoxon at risk, over crucial experiments, which are guaranteed to transform the Polydoxon, even if incrementally.

A second type of Polydoxon-neutral activities include performing computations within a speculative theoretical framework that makes no contribution to its prospects of discovery. For example, getting precisely correct the chargino pair production cross section at a high energy $e^+e^-$ collider will not change the prospects of finding the chargino. It would have a very small effect on the ruled out parameter space in the future once the collider begins its run, but theory investment for such incremental updating of the supersymmetric parameter space is not worth the minuscule transformation of the Polydoxon. In contrast, computations that point out qualitatively new discovery options, such as new signatures that would not have been looked at without it being pointed out, do hold good promise for Polydoxon transformation.

A related third type of Polydoxon-neutral activity is doing precision calculations within a theory that yields results with significantly lower error than any reasonably contemplated experiment could give. Or performing higher-loop computations of processes in speculative theories, when such computations have no impact on discovering unique features of or ruling out theories.  Or, similarly, doing higher order calculations in a theory that will never improve the comparison with experiment. An example of that would be three-loop electroweak computations of $b\to s\gamma$ flavor-changing neutral current quark decays. Such a computation would take years, and would have no value because experimental extraction is too uncertain to take advantage of the calculation that itself would have problematic interpretation for comparing theory to experiment.

A fourth Polydoxon-neutral activity is positing a future with new data that points to dramatic new discovery of a new particle, for example, and then determining how the Polydoxon can be transformed by additional analysis. These  speculative future Polydoxon transformations exercises are unlikely to be rewarded because there are an infinite number of future Polydoxons that could happen. One can imagine how contemplating  speculative future Polydoxons through consideration of high-credence theories in the current Polydoxon could be enticing~\cite{Arkani-Hamed:2005qjb} based on this high credence motivation, but they drain effort away from the more likely rewarded effort of finding a way to transform the current Polydoxon in front of us.

A fifth type of Polydoxon-neutral activities are ones that are entirely soft interpretation based. An example of this is discussing the subtleties of many worlds versus Copenhagen interpretations of quantum mechanics. There is currently little prospect of empirical differentiation between them, and thus such engagement is unlikely to be rewarded in the physics community, but perhaps in other communities, such as philosophy of science.

Finally, we emphasize here that although Polydoxon-neutral activities, as described above, are unlikely to be rewarded at a high level in physics, it does not necessarily mean that such activities are to be eschewed. Some of these activities can lead to progress, or depth of knowledge of researchers, or service to future generations, or honing experimental skills in ready-made environments, or building better credence models, etc. In other words, we seek not here to be prejudicial about such activities, but only remark that the data on highly rewarded activities suggests that Polydoxon-neutral work is less recognized.

\xsubsection{Non-empirical assessments and the Polydoxon}

One might add to the list of Polydoxon-neutral activities any discussions of non-empirical qualities of theories.  Examples of this type would be ranking theories based on naturalness, simplicity, or any other subjective criteria. Such considerations generally do not change membership of the Polydoxon, as they do not change what theories are empirically viable. For example, noting that an empirically viable theory is highly finetuned does not change the fact that it is empirically viable and remains a member of the Polydoxon.

However, non-empirical evaluations can change the credences of theories or the utility scores of theories when evaluating theories within the Polydoxon. This can have an effect on pursuit worthiness of activities focused on theories with higher credence or higher utility scores. If we allow that non-empirical assessments of theories can change the meta-structure (i.e., rankings of member theories) of the Polydoxon, by radical realignments of our credences or utilities with respect to them, then they can be thought of as judgments to set priorities in pursuit of transforming the Polydoxon.

\xsubsection{Tenure as first major award}

For many early career physicists the first and perhaps most important award they receive is gaining tenure at a top research university. There are many criteria for obtaining tenure, including teaching and service, but the key criteria, and the ones that are of interest to us here, are the research accomplishments. There are thousands of physicists with tenure, and so it is not an especially exclusive award. Nevertheless, it is a very significant development in a researcher's career and it is useful to briefly discuss how transformations of the Polydoxon may be relevant.

One of the best ways to determine what top research universities are looking for is to inspect the letters that solicit external reviews from top researchers at other institutions. All such letters include language asking reviewers to comment on the candidate's ``depth, originality, ... impact ... [and] future potential''~\cite{Brown:2024}, to quote one such typical wording.

As applied to physicists, these external review requests are somewhat vague. It does not explicitly make requirements for a physicist to be relevant to the transformation of the set of empirically viable theories. Nevertheless, there is an interpretation that is consistent with that understanding. For example, we can say that what the solicitation to external reviewers is equivalently asking is whether the researcher has creatively expanded or tested the Polydoxon (``originality''). Have they transformed the Polydoxon in significant ways recognized by experts (``depth'' and ``impact''), or does their work include substantive enabling moves that strategically position themselves or  the field for significant future transformations of the Polydoxon (``potential'')? These are the accomplishments valued in receiving the award of secure employment to continue one's scientific work, and there is a clear connection that can be made with significant transformations of the Polydoxon.

%%%%%%%%%%%%%%%%%%%%%%%%%%%%%%%%%%%%%%%%%%%%%%%%%%%%%%%
\section{Magnitude of Polydoxon transformations}

The classification of Polydoxon transformations into $P_+$, $P_-$, $P_R$, and $P_E$ is intentionally broad. One might therefore object that nearly all physics research activity can be described within this scheme, and that such a classification cannot, by itself, explain why some contributions are highly rewarded while others are not. However, we have already shown that there are significant activities that take place within physics that are neutral to the Polydoxon and in some cases corrupting to the Polydoxon. Thus, the classification of Polydoxon transformations decisively do not account for all activities, and alone is useful to identify a (usually) necessary but not sufficient condition for highly rewarded activity.  Nevertheless, these considerations point to an important refinement in our descriptive theory of rewarded activities. The categories $P_+$, $P_-$, $P_R$, and $P_E$ specify the type of transformation effected on the Polydoxon; they do not, on their own, measure the magnitude or significance of that transformation.

Most research activity produces only incremental, local, or weakly consequential changes in the structure of empirically viable theory space. Indeed, as discussed above, a substantial fraction of activity is effectively Polydoxon-neutral, in the sense that it neither meaningfully alters the membership of the Polydoxon nor significantly reorganizes its structure or future prospects. By contrast, highly rewarded contributions are those that produce transformations of unusually large scope or strategic importance. The explanatory content of the present account therefore lies not merely in identifying the type of transformation, but in recognizing that rewards correspond to those transformations that are large, central, robust, or generative of future change.

To make this more precise, it is useful to distinguish several dimensions along which the magnitude of a Polydoxon transformation can vary:

{\it Extent of membership change.} Transformations differ in how much of the Polydoxon they affect. Additions may open large new regions of empirically viable theory space, while contractions may eliminate broad classes of theories or only small peripheral subsets. The discovery of the Higgs boson, for example, eliminated entire classes of alternative theories, whereas many experimental constraints rule out only narrow regions of parameter space.

{\it Centrality of impact.} Transformations that affect theories or structures that are central to ongoing research programs carry greater significance than those that affect peripheral or weakly connected regions. Eliminating or restructuring a highly connected or widely studied class of theories has a larger effect on the field than modifying an isolated niche. This is particularly relevant for transformations that involve theories given high credence within the community.

{\it Depth of reconfiguration.} For $P_R$ transformations, the magnitude is tied to how profoundly the relationships within the Polydoxon are reorganized. Insights that reveal deep structural connections, such as surprising dualities, universality under renormalization group flow, or newly discovered common underlying principles, have a qualitatively different impact from minor reformulations or clarifications.

{\it Future leverage.} Particularly for $P_E$ transformations, the magnitude is determined by the extent to which a contribution enhances the field's capacity to generate subsequent transformations. New experimental techniques, theoretical tools, or observational windows may have little immediate effect on the Polydoxon, yet dramatically increase the likelihood of future $P_+$, $P_-$, or $P_R$ transformations. Likewise, there are some less consequential $P_E$ transformations that are immediately impacting, but without centrality of impact.

These dimensions are not independent, and in practice highly rewarded contributions tend to score highly along several of them simultaneously. A radical contraction of central theory space that is robust and experimentally decisive represents a paradigmatic high-impact $P_-$ transformation. Likewise, a theoretical advance that reveals deep structural unity while enabling new lines of inquiry constitutes a high-impact $P_R$ transformation. Enabling moves of the highest significance are those that open qualitatively new domains of empirical or theoretical access, thereby raising the expected rate and scale of future transformations.

This refinement is directly reflected in the historical cases discussed in Section 4. The discovery of the Higgs boson is highly rewarded not merely because it is a $P_-$ transformation, but because it constitutes a large, central, and decisive contraction of theory space. The development of the renormalization group is not simply a $P_R$ transformation, but one of exceptional depth and future leverage, reorganizing the structure of quantum field theory and enabling entire research programs. The detection of gravitational waves is a $P_E$ transformation whose significance lies in its extraordinary future leverage, opening a new observational window with far-reaching implications for early universe cosmology and strong-gravity phenomena. In each case, it is the magnitude and strategic importance of the transformation across the dimensions identified above rather than its categorical type alone, that explains its high level of recognition.

Another area where these considerations are important is in the activity of model building, as it is often called. Low impact model building is the creation of theories that solve no outstanding conceptual conundrum nor do they respond to any data pressures. Adding new particles that answer no pressing questions, for example, is a very low impact activity and would constitute a small, incremental addition to the Polydoxon ($\delta P_+$) that is unlikely to be rewarded well within the community.

It is therefore not sufficient, for the purposes of understanding scientific reward, to note that a contribution falls into one of the categories $P_+$, $P_-$, $P_R$, or $P_E$. Rather, the key explanatory variable is the magnitude and strategic significance of the transformation within that category. In this sense, the present framework provides a first-order taxonomy of reward-relevant scientific activity, while the distinction between incremental and exceptional transformations supplies the second-order structure needed to account for the distribution of major scientific rewards.

%%%%%%%%%%%%%%%%%%%%%%%%%%%%%%%%%%%%%%%%%%%%%%%%%%%%%%%
\section{Conclusions}

We have presented here a descriptive account of physics advancement centered on the Polydoxon, which is the structured set of current empirically viable theories. One of our key results is that scientific rewards come largely through direct substantive transformations of the Polydoxon or significant enabling moves. The transformations add to ($P_+$), subtract from ($P_-$), or reconfigure ($P_R$) the Polydoxon. Enabling moves ($P_E)$ are ones that strategically position the field to make transformations of the Polydoxon.

We have made the claim that the Nobel Prize in physics represents the highest award in physics, and the vast majority of the Nobel Prizes have gone to activities that have clearly resulted in substantial transformations of the Polydoxon as we have defined it. Other major awards, such as the Sakurai Prize, the Panofsky Prize, and the APS Medal of the American Physical Society, were also used to illustrate the thesis. Research activities and results that fall short of winning major awards either transform the Polydoxon incrementally or not at all. In a previous section we listed numerous activities that physicists regularly engage in that are neutral to Polydoxon transformations, and therefore ultimately not rewarded. We have also discussed the various measures of impact for Polydoxon transformations that include the scope of membership change, centrality of impact, depth of reconfiguration, and extent of future leverage.

Although we have not shown it in this article, we believe that a normative theory of scientific progress can be formulated through the lens of this account. One key gateway in that argument would be to establish that major scientific awards constitute progress. As is well known, there is much to contemplate regarding unfairness and deficiencies that can occur with any reward system. Furthermore, there is much to contemplate regarding what ``progress'' means. These are not to be taken lightly, and so we have not attempted to prove the normative turn here that scientific progress takes place through activities primarily devoted to transforming the Polydoxon. 

What can be said, however, is that the principal pathways to the highest recognition in physics are contributions that transform the Polydoxon, contributions that render such transformations effectively inevitable (enabling moves), and, more rarely, engineering achievements with immediate and substantial societal impact.

%%%%%%%%%%%%%%%%%%%%%%%%%%%%%%%%%%%%%%%%%%%%%%%%%%%%%%%
%%%%%%%%%%%%%%%%%%%%%%%%%%%%%%%%%%%%%%%%%%%%%%%%%%%%%%%
%\vfill\eject
%%%%%%%%%%%%%
%%%%%%%%%%%%%%%%%%%%%%%%%%%%%%%%%%%%%%%%%%%%%%%%%%%%%%%
%%%%%%%%%%%%%%%%%%%%%%%%%%%%%%%%%%%%%%%%%%%%%%%%%%%%%%%
%%%%%%%%%%%%%%%%%%%%%%%%%%%%%%%%%%%%%%%%%%%%%%%%%%%%%%%
\bigskip
\noindent
{\it Acknowledgements:} Support from Leinweber Institute for Theoretical Physics is gratefully acknowledged. I gratefully acknowledge discussions on these topics with G.\ Kane, R.\ Akhoury, and S.\ Martin.

\medskip
\noindent
{\it Generative AI Usage Statement:} During preparation of this manuscript, the author used ChatGPT-5.5 as an aid in manuscript preparation. The tool was used to suggest wording for the abstract, introduction, section transitions, and sentence-level smoothing after substantive draft material had been prepared by the author. It was also used to assist in preparing an initial appendix table with preliminary category assignments and to suggest some potentially relevant references. The author independently reviewed, edited, corrected, and verified all AI-assisted material, including all classifications and references. The conceptual framework, scholarly interpretation, argumentation, final text, and final judgments are the author's own. The author takes full responsibility for the accuracy, integrity, and content of the article.

\section*{Appendix: Nobel Prizes in Physics by Decade}

Below are all the Nobel Prizes in Physics awarded to date. The date awarded, individuals awarded, and a brief description of the research is given. In the last column is the ``code'' for what category of research was impacted with respect to transformation of the Polydoxon:
$P_+$ (expansion), $P_-$ (contraction), $P_R$ (reconfiguring), and $P_E$ (enabling move). The ``Eng'' code indicates the award was for highly impactful engineering work.

While the classification of Polydoxon transformations into $P_+$, $P_-$, $P_R$, and $P_E$ is conceptually clear at the level of ideal types, individual prize-winning contributions often exhibit features of more than one category. For example, a theoretical advance may both reorganize the structure of existing theory space and enable future empirical tests, or an experimental discovery may simultaneously contract parts of the Polydoxon while inaugurating a new experimental regime. In such cases, the classification adopted here assigns a single code corresponding to the primary mode of impact, defined as the aspect of the contribution that best explains its historical significance and recognition by the community. This primary mode is identified by asking whether the work is most fundamentally remembered as expanding the space of empirically viable theories ($P_+$), contracting it through decisive empirical discrimination ($P_-$), reconfiguring its internal structure and relations ($P_R$), or strategically positioning  the field for future transformations ($P_E$, enabling move). This convention preserves the analytical clarity of the classification while acknowledging that many of the most consequential contributions derive their importance from a combination of these effects. In several cases, contributions plausibly fall into multiple categories; the assigned code reflects the dominant mode of impact, though secondary effects may be significant.

\section*{1900s}
\small
\begin{tabularx}{\textwidth}{@{}C{0.95cm}Y Y C{1.65cm}@{}}
\toprule
\textbf{Year} & \textbf{Laureate(s)} & \textbf{Brief description} & \textbf{Code} \\
\midrule
1901 & Wilhelm Conrad R\"ontgen & Discovery of X-rays & $P_+$ \\
1902 & Hendrik A. Lorentz; Pieter Zeeman & Research on the influence of magnetism on radiation phenomena & $P_-$ \\
1903 & Henri Becquerel & Discovery of spontaneous radioactivity & $P_+$ \\
1903 & Pierre Curie; Marie Curie & Research on radiation phenomena & $P_+$ \\
1904 & Lord Rayleigh & Gas densities and discovery of argon & $P_+$ \\
1905 & Philipp Lenard & Work on cathode rays & $P_+$ \\
1906 & J.J. Thomson & Theoretical and experimental investigations on conduction of electricity by gases & $P_+$ \\
1907 & Albert A. Michelson & Optical precision instruments and spectroscopic/metrological investigations & $P_E$ \\
1908 & Gabriel Lippmann & Method of reproducing colours photographically by interference & Eng \\
1909 & Guglielmo Marconi; Ferdinand Braun & Development of wireless telegraphy & Eng \\
\bottomrule
\end{tabularx}
\normalsize

\section*{1910s}
\small
\begin{tabularx}{\textwidth}{@{}C{0.95cm}Y Y C{1.65cm}@{}}
\toprule
\textbf{Year} & \textbf{Laureate(s)} & \textbf{Brief description} & \textbf{Code} \\
\midrule
1910 & Johannes Diderik van der Waals & Equation of state for gases and liquids & $P_+$ \\
1911 & Wilhelm Wien & Laws governing heat radiation & $P_+$ \\
1912 & Gustaf Dal\'en & Automatic regulators for lighthouses and buoys & Eng \\
1913 & Heike Kamerlingh Onnes & Low-temperature investigations leading to liquid helium & $P_+$ \\
1914 & Max von Laue & Discovery of X-ray diffraction by crystals & $P_+$ \\
1915 & William Bragg; Lawrence Bragg & Analysis of crystal structure by means of X-rays & $P_E$ \\
1916 &  $\text{---}$ & No Nobel Prize awarded this year & N/A \\
1917 & Charles Glover Barkla & Discovery of characteristic X-radiation of the elements & $P_+$ \\
1918 & Max Planck & Discovery of energy quanta & $P_+$ \\
1919 & Johannes Stark & Doppler effect in canal rays and splitting of spectral lines in electric fields & $P_+$ \\
\bottomrule
\end{tabularx}
\normalsize

\section*{1920s}
\small
\begin{tabularx}{\textwidth}{@{}C{0.95cm}Y Y C{1.65cm}@{}}
\toprule
\textbf{Year} & \textbf{Laureate(s)} & \textbf{Brief description} & \textbf{Code} \\
\midrule
1920 & Charles Edouard Guillaume & Precision measurements via anomalies in nickel steel alloys & $P_E$ \\
1921 & Albert Einstein & Services to theoretical physics, especially the law of the photoelectric effect & $P_+$ \\
1922 & Niels Bohr & Investigation of atomic structure and radiation & $P_+$ \\
1923 & Robert A. Millikan & Elementary charge and the photoelectric effect & $P_-$ \\
1924 & Manne Siegbahn & Discoveries and research in X-ray spectroscopy & $P_E$ \\
1925 & James Franck; Gustav Hertz & Laws governing electron impact on atoms & $P_-$ \\
1926 & Jean Baptiste Perrin & Work on the discontinuous structure of matter and sedimentation equilibrium & $P_-$ \\
1927 & Arthur H. Compton & Discovery of the Compton effect & $P_-$ \\
1927 & C.T.R. Wilson & Method of making charged-particle paths visible by condensation of vapour & $P_E$ \\
1928 & Owen Willans Richardson & Thermionic phenomenon and the Richardson law & $P_R$ \\
1929 & Louis de Broglie & Discovery of the wave nature of electrons & $P_+$ \\
\bottomrule
\end{tabularx}
\normalsize

\section*{1930s}
\small
\begin{tabularx}{\textwidth}{@{}C{0.95cm}Y Y C{1.65cm}@{}}
\toprule
\textbf{Year} & \textbf{Laureate(s)} & \textbf{Brief description} & \textbf{Code} \\
\midrule
1930 & Chandrasekhara Venkata Raman & Discovery of the Raman effect / work on light scattering & $P_+$ \\
1931 & $\text{---}$ & No Nobel Prize awarded this year & N/A \\
1932 & Werner Heisenberg & Creation of quantum mechanics & $P_+$ \\
1933 & Erwin Schr\"odinger; Paul A.M. Dirac & Discovery of new productive forms of atomic theory & $P_+$ \\
1934 & $\text{---}$ & No Nobel Prize awarded this year & N/A \\
1935 & James Chadwick & Discovery of the neutron & $P_+$ \\
1936 & Victor F. Hess & Discovery of cosmic radiation & $P_+$ \\
1936 & Carl D. Anderson & Discovery of the positron & $P_+$ \\
1937 & Clinton Davisson; George Paget Thomson & Experimental discovery of electron diffraction by crystals & $P_-$ \\
1938 & Enrico Fermi & New radioactive elements from neutron irradiation; slow-neutron nuclear reactions & $P_+$ \\
1939 & Ernest Lawrence & Invention and development of the cyclotron & $P_E$ \\
\bottomrule
\end{tabularx}
\normalsize

\section*{1940s}
\small
\begin{tabularx}{\textwidth}{@{}C{0.95cm}Y Y C{1.65cm}@{}}
\toprule
\textbf{Year} & \textbf{Laureate(s)} & \textbf{Brief description} & \textbf{Code} \\
\midrule
1940 & $\text{---}$ & No Nobel Prize awarded this year & N/A \\
1941 & $\text{---}$ & No Nobel Prize awarded this year & N/A \\
1942 & $\text{---}$ & No Nobel Prize awarded this year & N/A \\
1943 & Otto Stern & Molecular ray method and discovery of the proton magnetic moment & $P_E$ \\
1944 & Isidor Isaac Rabi & Resonance method for recording magnetic properties of atomic nuclei & $P_E$ \\
1945 & Wolfgang Pauli & Discovery of the exclusion principle & $P_R$ \\
1946 & Percy W. Bridgman & Apparatus for extremely high pressures and discoveries therewith & $P_E$ \\
1947 & Edward V. Appleton & Discovery of the Appleton layer in the upper atmosphere & $P_+$ \\
1948 & Patrick M.S. Blackett & Wilson cloud chamber method and discoveries in nuclear physics and cosmic radiation & $P_E$ \\
1949 & Hideki Yukawa & Prediction of mesons from theoretical work on nuclear forces & $P_+$ \\
\bottomrule
\end{tabularx}
\normalsize

\section*{1950s}
\small
\begin{tabularx}{\textwidth}{@{}C{0.95cm}Y Y C{1.65cm}@{}}
\toprule
\textbf{Year} & \textbf{Laureate(s)} & \textbf{Brief description} & \textbf{Code} \\
\midrule
1950 & Cecil Powell & Photographic method for nuclear processes and meson discoveries & $P_E$ \\
1951 & John Cockcroft; Ernest T.S. Walton & Artificial transmutation of atomic nuclei with accelerated particles & $P_E$ \\
1952 & Felix Bloch; E. M. Purcell & Methods for nuclear magnetic precision measurements & $P_E$ \\
1953 & Frits Zernike & Phase contrast method and microscope & $P_E$ \\
1954 & Max Born & Statistical interpretation of the wavefunction in quantum mechanics & $P_R$ \\
1954 & Walther Bothe & Coincidence method and discoveries made with it & $P_E$ \\
1955 & Willis E. Lamb & Discoveries concerning the fine structure of the hydrogen spectrum & $P_-$ \\
1955 & Polykarp Kusch & Precision determination of the electron magnetic moment & $P_-$ \\
1956 & William B. Shockley; John Bardeen; Walter H. Brattain & Research on semiconductors and discovery of the transistor effect & Eng \\
1957 & Chen Ning Yang; Tsung-Dao Lee & Investigation of parity laws leading to major discoveries in particle physics & $P_R$ \\
1958 & Pavel A. Cherenkov; I.M. Frank; Igor Y. Tamm & Discovery and interpretation of the Cherenkov effect & $P_R$ \\
1959 & Emilio Segr\`e; Owen Chamberlain & Discovery of the antiproton & $P_-$ \\
\bottomrule
\end{tabularx}
\normalsize

\section*{1960s}
\small
\begin{tabularx}{\textwidth}{@{}C{0.95cm}Y Y C{1.65cm}@{}}
\toprule
\textbf{Year} & \textbf{Laureate(s)} & \textbf{Brief description} & \textbf{Code} \\
\midrule
1960 & Donald A. Glaser & Invention of the bubble chamber & $P_E$ \\
1961 & Robert Hofstadter & Electron scattering studies revealing nucleon structure & $P_-$ \\
1961 & Rudolf M\"ossbauer & Discovery of the M\"ossbauer effect & $P_+$ \\
1962 & Lev Landau & Pioneering theories for condensed matter, especially liquid helium & $P_+$ \\
1963 & Eugene Wigner & Symmetry principles in nuclei and elementary particles & $P_R$ \\
1963 & Maria Goeppert Mayer; J. Hans D. Jensen & Discovery of nuclear shell structure & $P_+$ \\
1964 & Charles H. Townes; Nicolay G. Basov; Aleksandr M. Prokhorov & Quantum electronics leading to masers and lasers & $P_E$ \\
1965 & Sin-Itiro Tomonaga; Julian Schwinger; Richard P. Feynman & Fundamental work in quantum electrodynamics & $P_+$ \\
1966 & Alfred Kastler & Optical methods for studying Hertzian resonances in atoms & $P_E$ \\
1967 & Hans Bethe & Theory of nuclear reactions, especially stellar energy production & $P_R$ \\
1968 & Luis Alvarez & Bubble-chamber techniques and discovery of many resonance states & $P_E$ \\
1969 & Murray Gell-Mann & Classification of elementary particles and their interactions & $P_+$ \\
\bottomrule
\end{tabularx}
\normalsize

\section*{1970s}
\small
\begin{tabularx}{\textwidth}{@{}C{0.95cm}Y Y C{1.65cm}@{}}
\toprule
\textbf{Year} & \textbf{Laureate(s)} & \textbf{Brief description} & \textbf{Code} \\
\midrule
1970 & Hannes Alfv\'en & Fundamental work in magnetohydrodynamics & $P_R$ \\
1970 & Louis N\'eel & Fundamental work on antiferromagnetism and ferrimagnetism & $P_R$ \\
1971 & Dennis Gabor & Invention and development of holography & $P_E$ \\
1972 & John Bardeen; Leon N. Cooper; Robert Schrieffer & BCS theory of superconductivity & $P_+$ \\
1973 & Leo Esaki; Ivar Giaever & Experimental discoveries of tunneling phenomena in semiconductors and superconductors & $P_+$ \\
1973 & Brian D. Josephson & Theoretical prediction of the Josephson effect & $P_+$ \\
1974 & Martin Ryle & Observations and inventions in radio astrophysics, especially aperture synthesis & $P_E$ \\
1974 & Antony Hewish & Decisive role in the discovery of pulsars & $P_+$ \\
1975 & Aage N. Bohr; Ben R. Mottelson; James Rainwater & Theory of collective and particle motion in atomic nuclei & $P_R$ \\
1976 & Burton Richter; Samuel C.C. Ting & Discovery of a heavy elementary particle of a new kind (J/$\psi$) & $P_+$ \\
1977 & Philip W. Anderson; Sir Nevill F. Mott; John H. Van Vleck & Theoretical investigations of magnetic and disordered systems & $P_R$ \\
1978 & Pyotr Kapitsa & Basic inventions and discoveries in low-temperature physics & $P_E$ \\
1978 & Arno Penzias; Robert Woodrow Wilson & Discovery of the cosmic microwave background radiation & $P_+$ \\
1979 & Sheldon Glashow; Abdus Salam; Steven Weinberg & Theory unifying weak and electromagnetic interactions & $P_+$ \\
\bottomrule
\end{tabularx}
\normalsize

\section*{1980s}
\small
\begin{tabularx}{\textwidth}{@{}C{0.95cm}Y Y C{1.65cm}@{}}
\toprule
\textbf{Year} & \textbf{Laureate(s)} & \textbf{Brief description} & \textbf{Code} \\
\midrule
1980 & James Cronin; Val Fitch & Discovery of CP violation in neutral K-meson decay & $P_+$ \\
1981 & Nicolaas Bloembergen; Arthur L. Schawlow & Development of laser spectroscopy & $P_E$ \\
1981 & Kai M. Siegbahn & Development of high-resolution electron spectroscopy & $P_E$ \\
1982 & Kenneth G. Wilson & Theory for critical phenomena in phase transitions & $P_R$ \\
1983 & Subrahmanyan Chandrasekhar & Theoretical studies of stellar structure and evolution & $P_R$ \\
1983 & William A. Fowler & Theoretical and experimental studies of nuclear reactions forming the chemical elements & $P_R$ \\
1984 & Carlo Rubbia; Simon van der Meer & Contributions leading to the discovery of the W and Z bosons & $P_E$ \\
1985 & Klaus von Klitzing & Discovery of the quantized Hall effect & $P_+$ \\
1986 & Ernst Ruska & Electron optics and the first electron microscope & $P_E$ \\
1986 & Gerd Binnig; Heinrich Rohrer & Design of the scanning tunneling microscope & $P_E$ \\
1987 & J. Georg Bednorz; K. Alex M\"uller & Discovery of superconductivity in ceramic materials & $P_+$ \\
1988 & Leon M. Lederman; Melvin Schwartz; Jack Steinberger & Neutrino beam method and discovery of the muon neutrino & $P_-$ \\
1989 & Norman F. Ramsey & Separated oscillatory fields method and its use in atomic clocks & $P_E$ \\
1989 & Hans G. Dehmelt; Wolfgang Paul & Development of the ion trap technique & $P_E$ \\
\bottomrule
\end{tabularx}
\normalsize

\section*{1990s}
\small
\begin{tabularx}{\textwidth}{@{}C{0.95cm}Y Y C{1.65cm}@{}}
\toprule
\textbf{Year} & \textbf{Laureate(s)} & \textbf{Brief description} & \textbf{Code} \\
\midrule
1990 & Jerome I. Friedman; Henry W. Kendall; Richard E. Taylor & Deep inelastic scattering, crucial for development of the quark model & $P_-$ \\
1991 & Pierre-Gilles de Gennes & Generalizing methods for order phenomena to liquid crystals and polymers & $P_R$ \\
1992 & Georges Charpak & Invention and development of particle detectors, especially the multiwire proportional chamber & $P_E$ \\
1993 & Russell A. Hulse; Joseph H. Taylor Jr. & Discovery of a new type of pulsar opening new studies of gravitation & $P_-$ \\
1994 & Bertram N. Brockhouse & Development of neutron spectroscopy & $P_E$ \\
1994 & Clifford G. Shull & Development of neutron diffraction & $P_E$ \\
1995 & Martin L. Perl & Discovery of the tau lepton & $P_+$ \\
1995 & Frederick Reines & Detection of the neutrino & $P_-$ \\
1996 & David M. Lee; Douglas D. Osheroff; Robert C. Richardson & Discovery of superfluidity in helium-3 & $P_+$ \\
1997 & Steven Chu; Claude Cohen-Tannoudji; William D. Phillips & Methods to cool and trap atoms with laser light & $P_E$ \\
1998 & Robert B. Laughlin; Horst L. St\"ormer; Daniel C. Tsui & Discovery of a new quantum fluid with fractionally charged excitations & $P_+$ \\
1999 & Gerardus 't Hooft; Martinus J.G. Veltman & Elucidating the quantum structure of electroweak interactions & $P_R$ \\
\bottomrule
\end{tabularx}
\normalsize

\section*{2000s}
\small
\begin{tabularx}{\textwidth}{@{}C{0.95cm}Y Y C{1.65cm}@{}}
\toprule
\textbf{Year} & \textbf{Laureate(s)} & \textbf{Brief description} & \textbf{Code} \\
\midrule
2000 & Zhores Alferov; Herbert Kroemer & Semiconductor heterostructures for high-speed and optoelectronics & Eng \\
2000 & Jack Kilby & Invention of the integrated circuit & Eng \\
2001 & Eric Cornell; Wolfgang Ketterle; Carl Wieman & Achievement of Bose-Einstein condensation in dilute gases and early studies of condensates & $P_E$ \\
2002 & Raymond Davis Jr.; Masatoshi Koshiba & Detection of cosmic neutrinos & $P_E$ \\
2002 & Riccardo Giacconi & Discovery of cosmic X-ray sources & $P_+$ \\
2003 & Alexei Abrikosov; Vitaly L. Ginzburg; Anthony J. Leggett & Pioneering contributions to the theory of superconductors and superfluids & $P_R$ \\
2004 & David J. Gross; H. David Politzer; Frank Wilczek & Discovery of asymptotic freedom in the strong interaction & $P_R$ \\
2005 & Roy J. Glauber & Contribution to the quantum theory of optical coherence & $P_R$ \\
2005 & John L. Hall; Theodor W. H\"ansch & Laser-based precision spectroscopy and the optical frequency comb & $P_E$ \\
2006 & John C. Mather; George F. Smoot & Discovery of the blackbody form and anisotropy of the cosmic microwave background & $P_-$ \\
2007 & Albert Fert; Peter Gr\"unberg & Discovery of giant magnetoresistance & $P_+$ \\
2008 & Yoichiro Nambu & Discovery of the mechanism of spontaneous broken symmetry in subatomic physics & $P_R$ \\
2008 & Makoto Kobayashi; Toshihide Maskawa & Discovery of the origin of broken symmetry predicting at least three quark families & $P_+$ \\
2009 & Charles K. Kao & Transmission of light in fibers for optical communication & Eng \\
2009 & Willard S. Boyle; George E. Smith & Invention of the CCD image sensor & Eng \\
\bottomrule
\end{tabularx}
\normalsize

\section*{2010s}
\small
\begin{tabularx}{\textwidth}{@{}C{0.95cm}Y Y C{1.65cm}@{}}
\toprule
\textbf{Year} & \textbf{Laureate(s)} & \textbf{Brief description} & \textbf{Code} \\
\midrule
2010 & Andre Geim; Konstantin Novoselov & Groundbreaking experiments on graphene & $P_+$ \\
2011 & Saul Perlmutter; Brian P. Schmidt; Adam G. Riess & Discovery of the accelerating expansion of the universe through distant supernovae & $P_+$ \\
2012 & Serge Haroche; David J. Wineland & Experimental methods for measuring and manipulating individual quantum systems & $P_E$ \\
2013 & Fran\c{c}ois Englert; Peter Higgs & Theoretical discovery of the Higgs mechanism, later confirmed experimentally & $P_+$ \\
2014 & Isamu Akasaki; Hiroshi Amano; Shuji Nakamura & Invention of efficient blue light-emitting diodes & Eng \\
2015 & Takaaki Kajita; Arthur B. McDonald & Discovery of neutrino oscillations, showing that neutrinos have mass & $P_-$ \\
2016 & David J. Thouless; F. Duncan M. Haldane; J. Michael Kosterlitz & Theoretical discoveries of topological phase transitions and topological phases of matter & $P_R$ \\
2017 & Rainer Weiss; Barry C. Barish; Kip S. Thorne & Decisive contributions to LIGO and the observation of gravitational waves & $P_E$ \\
2018 & Arthur Ashkin & Optical tweezers and their application to biological systems & $P_E$ \\
2018 & G\'erard Mourou; Donna Strickland & Method of generating high-intensity, ultra-short optical pulses & $P_E$ \\
2019 & James Peebles & Theoretical discoveries in physical cosmology & $P_R$ \\
2019 & Michel Mayor; Didier Queloz & Discovery of an exoplanet orbiting a solar-type star & $P_+$ \\
\bottomrule
\end{tabularx}
\normalsize

\section*{2020s}
\small
\begin{tabularx}{\textwidth}{@{}C{0.95cm}Y Y C{1.65cm}@{}}
\toprule
\textbf{Year} & \textbf{Laureate(s)} & \textbf{Brief description} & \textbf{Code} \\
\midrule
2020 & Roger Penrose & Discovery that black-hole formation is a robust prediction of general relativity & $P_R$ \\
2020 & Reinhard Genzel; Andrea Ghez & Discovery of a supermassive compact object at the centre of the Milky Way & $P_-$ \\
2021 & Syukuro Manabe; Klaus Hasselmann & Physical modelling of Earth's climate, quantifying variability, and predicting global warming & $P_+$ \\
2021 & Giorgio Parisi & Discovery of the interplay of disorder and fluctuations in physical systems & $P_R$ \\
2022 & Alain Aspect; John Clauser; Anton Zeilinger & Experiments with entangled photons, Bell-inequality violations, and quantum information science & $P_-$ \\
2023 & Pierre Agostini; Ferenc Krausz; Anne L'Huillier & Experimental methods generating attosecond light pulses to study electron dynamics & $P_E$ \\
2024 & John J. Hopfield; Geoffrey Hinton & Foundational discoveries and inventions enabling machine learning with artificial neural networks & $P_E$ \\
2025 & John Clarke; Michel H. Devoret; John M. Martinis & Discovery of macroscopic quantum mechanical tunnelling and energy quantisation in an electric circuit & $P_E$ \\
\bottomrule
\end{tabularx}
\normalsize

%%%%%%%%%%%%%%%%%%%%%%%%%%%%%%%%%%%%%%%%%%%%%%%%%%%%%%%
%%%%%%%%%%%%%%%%%%%%%%%%%%%%%%%%%%%%%%%%%%%%%%%%%%%%%%%


\begin{thebibliography}{99}

%\cite{LIGOScientific:2016aoc}
\bibitem{LIGOScientific:2016aoc}
B.~P.~Abbott \textit{et al.} [LIGO Scientific and Virgo],
``Observation of Gravitational Waves from a Binary Black Hole Merger,''
Phys. Rev. Lett. \textbf{116}, no.6, 061102 (2016)
doi:10.1103/PhysRevLett.116.061102
[arXiv:1602.03837 [gr-qc]].


\bibitem{APS:Medal}
American Physical Society, ``APS Medal for Exceptional Achievement in Research.'' \url{
https://www.aps.org/funding-recognition/prize/aps-medal} (accessed 28 March 2026)

\bibitem{Sakurai-Randall-Sundrum}
American Physical Society, ``Spring 2019 American Physical Society Prizes and Awards Announced.'' Press Release, 23 October 2018.
\url{https://www.aps.org/about/news/2018/10/spring-2019-prizes-awards-announced}

\bibitem{Dannie:APS}
American Physical Society, ``Dannie Heineman Prize for Mathematical Physics.'' https://www.aps.org/funding-recognition/prize/dannie-heineman (accessed April 2, 2026)


%\cite{Antoniadis:1998ig}
\bibitem{Antoniadis:1998ig}
I.~Antoniadis, N.~Arkani-Hamed, S.~Dimopoulos and G.~R.~Dvali,
``New dimensions at a millimeter to a Fermi and superstrings at a TeV,''
Phys. Lett. B \textbf{436}, 257-263 (1998)
doi:10.1016/S0370-2693(98)00860-0
[arXiv:hep-ph/9804398 [hep-ph]].
%4920 citations counted in INSPIRE as of 18 Apr 2026

%\cite{Arkani-Hamed:1998jmv}
\bibitem{Arkani-Hamed:1998jmv}
N.~Arkani-Hamed, S.~Dimopoulos and G.~R.~Dvali,
``The Hierarchy problem and new dimensions at a millimeter,''
Phys. Lett. B \textbf{429}, 263-272 (1998)
doi:10.1016/S0370-2693(98)00466-3
[arXiv:hep-ph/9803315 [hep-ph]].

%\cite{Arkani-Hamed:2005qjb}
\bibitem{Arkani-Hamed:2005qjb}
N.~Arkani-Hamed, G.~L.~Kane, J.~Thaler and L.~T.~Wang,
``Supersymmetry and the LHC inverse problem,''
JHEP \textbf{08}, 070 (2006)
doi:10.1088/1126-6708/2006/08/070
[arXiv:hep-ph/0512190 [hep-ph]].
%149 citations counted in INSPIRE as of 18 Apr 2026

%\cite{Arkani-Hamed:2000ijo}
\bibitem{Arkani-Hamed:2000ijo}
N.~Arkani-Hamed, M.~Porrati and L.~Randall,
``Holography and phenomenology,''
JHEP \textbf{08}, 017 (2001)
doi:10.1088/1126-6708/2001/08/017
[arXiv:hep-th/0012148 [hep-th]].


\bibitem{Aspect:1982}
A.~Aspect, R.~Dalibard, G.~Roger, ``Experiment Test of Bell's Inequalities Using Time-Varying Analyzers,'' {\it Phys.\ Rev.\ Lett.}\ \textbf{49}, 1804-1807 (1982).
DOI: https://doi.org/10.1103/PhysRevLett.49.1804


\bibitem{Balzer:1987} W. Balzer, C.U.~Moulines, and J.~Sneed, {\it An Architectonic for Science.} Springer-Nature, 1987.

%\cite{Baumann:2009ds}
\bibitem{Baumann:2009ds}
D.~Baumann,
``Inflation,''
doi:10.1142/9789814327183{\_}0010
[arXiv:0907.5424 [hep-th]].

%\cite{Bell:1964kc}
\bibitem{Bell:1964kc}
J.~S.~Bell,
``On the Einstein-Podolsky-Rosen paradox,''
Physics Physique Fizika \textbf{1}, 195-200 (1964)
doi:10.1103/PhysicsPhysiqueFizika.1.195

\bibitem{Berger:1972}
J.\ Berger, B.P.\ cohen, M.\ Zelditch, Jr.\ ``Status characteristics and social interaction.'' {\it American Sociological Review} vol.\ 37, no.\ 3, June 1972.
\url{https://www.jstor.org/stable/2093465}

%\cite{Bilenky:2012qb}
\bibitem{Bilenky:2012qb}
S.~M.~Bilenky,
``Neutrino. History of a unique particle,''
Eur. Phys. J. H \textbf{38}, 345-404 (2013)
doi:10.1140/epjh/e2012-20068-9
[arXiv:1210.3065 [hep-ph]].
%70 citations counted in INSPIRE as of 18 Apr 2026

\bibitem{Bourdieu:2004}
Pierre Bourdieu. {\it Science of Science and Reflexivity.} University of Chicago Press, 2004.


\bibitem{Branahl:2025}
Johannes Branahl, ``Stagnant Lakatosian Research Programmes.'' Eur.\ J.\ Phil.\ Sci.\ 15, 53 (2025). arXiv preprint arXiv:2404.18307, 2024.
\url{https://link.springer.com/article/10.1007/s13194-025-00677-x}

\bibitem{Brown:2024}
Brown University, ``Draft Department Letter for Soliciting External Reviews'', Handbook of Academic Administration, 1 August 2024. \url{https://dof.brown.edu/sites/default/files/Appendix%20B.%20Draft%20Department%20Letter%20for%20Soliciting%20External%20Reviews_0.pdf} 
(accessed March 21, 2026).

\bibitem{BreakthroughPrize}
Breakthrough Prize Board. ``Fundamental Physics Breakthrough Prize.'' 
\url{https://breakthroughprize.org/Prize/1} (accessed 2 April 2026)

\bibitem{Chen:2023}
Y.~Chen, J.~Ding, ``Exploitation and exploration: An analysis of the research pattern of Nobel laureates in Physics.'' {\it J.\ of Informatics.} vol.\ 17, issue 3, 101428, August 2023. \url{https://www.sciencedirect.com/science/article/abs/pii/S1751157723000536}

\bibitem{Clauser:1972}
S.~Freedman, J.F.~Clauser, ``Experimental Test of Local Hidden-Variable Theories,'' {\it Phys.\ Rev.\ Lett.}\ \textbf{28}, 938-941 (1972). DOI: https://doi.org/10.1103/PhysRevLett.28.938

%\cite{COBE:1992syq}
\bibitem{COBE:1992syq}
G.~F.~Smoot \textit{et al.} [COBE],
``Structure in the COBE differential microwave radiometer first year maps,''
Astrophys. J. Lett. \textbf{396}, L1-L5 (1992)
doi:10.1086/186504

\bibitem{ChapelHill:1957}
C\'ecile~DeWitt and Dean~Rickles (eds.)``The Role of Gravitation in Physics: Report from the 1957 Chapel Hill Conference'', communicated by J\"urgen Renn, Alexander Blum and Peter Damerow. Chapel Hill 1957, Max Planck Research Library for History and Development of Knowledge, Sources 5, Edition Open Access 2017.
\url{https://edition-open-sources.org/media/sources/5/Sources5.pdf}

\bibitem{DiracMedal}
``The Dirac Medal.'' ICTP Trieste.
\url{https://www.ictp.it/prize/dirac-medal} (accessed 2 April 2026)

\bibitem{Douven:2022}
I.~Douven, {\it The Art of Abduction}. Boston: MIT Press, 2022.


\bibitem{APS:2014}
D. Ehrenstein. ``Blue was the Hardest Color.'' Physics 7, 103, 2014. 
https://physics.aps.org/articles/v7/103 

\bibitem{Dawid:2013}
R.~Dawid, {\it String Theory and the Scientific Method}. Cambridge University Press, 2013.


%\cite{Dawid:2024hqa}
\bibitem{Dawid:2024hqa}
R.~Dawid and J.~D.~Wells,
``A bayesian model of credence in low energy supersymmetry,''
Synthese \textbf{206}, no.4, 173 (2025)
doi:10.1007/s11229-025-05252-8
[arXiv:2411.03232 [physics.hist-ph]].


\bibitem{vanFraassen:1980} 
B.~van Fraassen, {\it The Scientific Image.} Oxford University Press, 1980.

\bibitem{Frigg:2006} R.~Frigg. ``Scientific Representation and the Semantic View of Theories.'' {\it THEORIA. An International Journal for Theory, History and Foundations of Science, 21(1), 49-65, 2006.} \url{https://doi.org/10.1387/theoria.553}

\bibitem{Frigg:SEP:2006} R.~Frigg and S.~Hartmann, ``Models in Science'' {\it Stanford Enclyclopedia of Philosophy.} 27 February 2006. \url{https://plato.stanford.edu/entries/models-science/}

\bibitem{Giere:1990}
R.N.~Giere, {\it Explaining Science: A Cognitive Approach.} University of Chicago Press, 1990.

%\cite{Heijne:1980rt}
\bibitem{Heijne:1980rt}
E.~H.~M.~Heijne, L.~Hubbeling, B.~D.~Hyams, P.~Jarron, P.~Lazeyras, F.~Piuz, J.~C.~Vermeulen and A.~Wylie,
``A silicon surface barrier microstrip detector designed for high-energy physics,"
%``A SILICON SURFACE BARRIER MICROSTRIP DETECTOR DESIGNED FOR HIGH-ENERGY PHYSICS,''
Nucl. Instrum. Meth. \textbf{178}, 331-343 (1980)
doi:10.1016/0029-554X(80)90812-5

\bibitem{Horvath:2023}
J.E.~Horvath, ``Contemporary Cosmology from Lakatos' Viewpoint'' {\it Cosmos and History: The Journal of Natural and Social Philosophy}, vol.\ 19, no.\ 1 (2023). arXiv preprint arXiv:2309.15695, 2023. \url{https://cosmosandhistory.org/index.php/journal/article/view/1095/1736}

%\cite{Isidori:2023pyp}
\bibitem{Isidori:2023pyp}
G.~Isidori, F.~Wilsch and D.~Wyler,
``The standard model effective field theory at work,''
Rev. Mod. Phys. \textbf{96}, no.1, 1 (2024)
doi:10.1103/RevModPhys.96.015006
[arXiv:2303.16922 [hep-ph]].

\bibitem{Kuhn:1962}
Thomas~Kuhn. {\it The Structure of Scientific Revolutions.} Chicago: University of Chicago Press, 1962.

\bibitem{Lakatos:1978}
I.~Lakatos, ``History of Science and Its Rational Reconstructions.'' In {\it The Methodology of Scientific Research Programmes}, edited by John Worrall and Gregory Currie, 102?138. Cambridge: Cambridge University Press, 1978.

\bibitem{Lakatos:1978b}
I.~Lakatos, {\it The Methodology of Scientific Research Programmes: volume 1: Philosophical Papers, 1st ed.}\ eds.\ J. Worrall, G. Currie. Cambridge University Press, 1978.

\bibitem{Laudan:1978}
L.\ Laudan. {\it Progress and its Problems: Towards a Theory of Scientific Growth}. Los Angeles: University of California Press, 1978.


\bibitem{Lipton:2004}
P.~Lipton, {\it Inference to the Best Explanation}, 2nd ed. New York: Routledge, 2004.

\bibitem{Marshall:2017}
J.A.R.~Marshall, G.~Brown, A.N.~Radford, 
``Individual Confidence-Weighting and Group Decision-Making''
Trends in Ecology \& Evolution, vol.\ 32, issue 9, 636-645 (2017).
\url{https://www.sciencedirect.com/science/article/pii/S0169534717301520}


\bibitem{Maxwell:2012}
Nicholas~Maxwell, ``Popper, Kuhn, Lakatos and Aim-Oriented Empiricism.'' {\it Philosophia} 32 (1-4), p.181-239 (2005).  arXiv preprint arXiv:1208.5219, 2012.
\url{https://philarchive.org/rec/MAXPKL-3}

\bibitem{Morgan:2010} M.S.~Morgan and M.~Morrison, M. (eds.) (2010), {\it Models as Mediators} Cambridge University Press, 2010

\bibitem{NAS:members}
National Academies of Science, ``Nomination \& Election.'' 
\url{https://www.nasonline.org/membership/nomination-election/} (accessed 2 April 2026)

\bibitem{Nilesh:2018}
Barde~Nilesh and Bardapurkar Pranav. ``Statistical analysis of Nobel Prizes in Physics: from its inception till date.'' {\it J.\ Physical Studies} vol.\ 22, no.\ 3, 3002, 2018.
\url{https://physics.lnu.edu.ua/jps/2018/3/pdf/3002-8.pdf}


\bibitem{Nobel:will}
Afred~Nobel, ``Alfred Nobel's Will." Nobel Prize Foundation. \url{https://www.nobelpeaceprize.org/nobel-peace-prize/history/alfred-nobel-s-will} (accessed April 18, 2026)

\bibitem{NobelPrizes}
The Nobel Foundation, ``Nobel Prizes in Physics,'' \\ https://www.nobelprize.org/prizes/lists/all-nobel-prizes-in-physics/ (accessed 2 April 2026).

\bibitem{Nobel:nomination}
The Nobel Foundation, ``Nomination and selection of physics laureates.'' The Nobel Foundation. https://www.nobelprize.org/nomination/physics/ (accessed 28 March 2026).


\bibitem{Pais:1950}
A.~Pais and G.\ Uhlenbeck. ``On field theories with non-localized action.'' {\it Phys.\ Rev.\ } 79, 145 (1950). https://journals.aps.org/pr/abstract/10.1103/PhysRev.79.145


\bibitem{PanofskyPrizes}
``W.K.H. Panofsky Prize in Experimental Particle Physics'', American Physical Society, https://\url{www.aps.org/funding-recognition/prize/panofsky}    (accessed 2 April 2026)


\bibitem{Popper:2002a}
Karl~Popper. {\it The Logic of Scientific Discovery.} 2nd ed.\ New York, Routledge, 2002 [1934/1959].

\bibitem{Popper:2002b}
Karl~Popper. {\it Conjectures and Refutations: The Growth of Scientific Knowledge,} 2nd ed.\ New York: Routledge, 2002 [1963].

%\cite{Randall:1999ee}
\bibitem{Randall:1999ee}
L.~Randall and R.~Sundrum,
``A Large mass hierarchy from a small extra dimension,''
Phys. Rev. Lett. \textbf{83}, 3370-3373 (1999)
doi:10.1103/PhysRevLett.83.3370
[arXiv:hep-ph/9905221 [hep-ph]].

\bibitem{Rossiter:1993}
M.W.~Rossiter, ``The Matthew/Matilda Effect in Science.'' {\it Social Studies of Science} 23, no. 2, 325-341 (1993). \url{https://www.jstor.org/stable/i212628}

\bibitem{SakuraiPrizes}
``J. J. Sakurai Prize for Theoretical Particle Physics,'' American Physical Society,
https://www.aps.org/funding-recognition/prize/sakurai-prize (accessed 2 April 2026)

\bibitem{Sepetyl:2025}
D.~Sepetyl, ``Popper and Lakatos on what is distinctive about empirical science'' {\it Intl.\ Stud.\ Phil.\ Sci.}\ vol.\ 38, issue 1 (2025). \url{https://doi.org/10.1080/02698595.2025.2449809}

\bibitem{Suppe:1989} 
F.~Suppe, {\it The Semantic Conception of Theories and Scientific Realism}. University of Illinois Press, 1989.

\bibitem{Tindale:2019}
R.S.~Tindale and J.R.~Winget, ``Group Decision-Making''
{\it Oxford Research Encyclopedia of Psychology.} 26 March 2019.
\url{https://doi.org/10.1093/acrefore/9780190236557.013.262} (accessed 28 March 2026)

%\cite{Trimble:1973ca}
\bibitem{Trimble:1973ca}
V.~Trimble and F.~Reines,
``The solar neutrino problem - a progress(?) report,''
Rev. Mod. Phys. \textbf{45}, 1-5 (1973)
doi:10.1103/RevModPhys.45.1
%56 citations counted in INSPIRE as of 18 Apr 2026

\bibitem{Zeilinger:1998}
G.~Weihs, T.~Jennewein, C.~Simon, H.~Weinfurter, A.~Zeilinger, ``Violation of Bell's Inequality under Strict Einstein Locality Conditions,'' {\it Phys.\ Rev.\ Lett.}\ \textbf{81}, 5039-5042 (1998).
DOI: https://doi.org/10.1103/PhysRevLett.81.5039

\bibitem{Weisberg:2007}
M.~Weisberg, ``Three Kinds of Idealization,'' J.~Philosophy, vol.104, No.12, p.630-659 (December 2007). \url{https://www.jstor.org/stable/20620065}

%\cite{Wells:2018nwj}
\bibitem{Wells:2018nwj}
J.~D.~Wells,
``Beyond the hypothesis: Theory's role in the genesis, opposition, and pursuit of the Higgs boson,''
Stud. Hist. Phil. Sci. B \textbf{62}, 36-44 (2018)
doi:10.1016/j.shpsb.2017.05.004

%\cite{Wells:2019rwq}
\bibitem{Wells:2019rwq}
J.~D.~Wells,
``The Once and Present Standard Model of Elementary Particle Physics,'' in {\it Discovery Beyond the Standard Model of Elementary Particle Physics}, New York: Springer, 2020.
\url{https://link.springer.com/book/10.1007/978-3-030-38204-9}

%\cite{Witten:1995ex}
\bibitem{Witten:1995ex}
E.~Witten,
``String theory dynamics in various dimensions,''
Nucl. Phys. B \textbf{443}, 85-126 (1995)
doi:10.1016/0550-3213(95)00158-O
[arXiv:hep-th/9503124 [hep-th]].

\bibitem{Woodard:2015}
R.P.~Woodard, ``The Theorem of Ostrogradsky.'' Scholarpedia 10(8), 32243 (2015). arXiv:1506.02210.
\url{http://www.scholarpedia.org/article/Ostrogradsky%27s_theorem_on_Hamiltonian_instability}

%\cite{Yang:1954ek}
\bibitem{Yang:1954ek}
C.~N.~Yang and R.~L.~Mills,
``Conservation of Isotopic Spin and Isotopic Gauge Invariance,''
Phys. Rev. \textbf{96}, 191-195 (1954)
doi:10.1103/PhysRev.96.191

%\cite{Yu:2025rez}
\bibitem{Yu:2025rez}
T.~T.~Yu,
``2024 TASI Lectures: A Dark Matter Primer,''
[arXiv:2506.05234 [hep-ph]].

\bibitem{Zhou:2014}
Zhiwei Zhou, Rui Xing, Jing Liu, Feiyue Xing, 2014. ``Landmark papers written by the Nobelists in physics from 1901 to 2012: a bibliometric analysis of their citations and journals,
 {\it Scientometrics}, Springer vol. 100(2), pages 329-338, August 2014.
\url{https://ideas.repec.org/a/spr/scient/v100y2014i2d10.1007_s11192-014-1306-7.html}

\bibitem{Zinn-Justin:2021}
J.~Zinn-Justin. {\it Quantum Field Theory and Critical Phenomena}, 5th ed. Oxford University Press, 2021.

\bibitem{Zuckerman:1977}
Harriet~Zuckerman, {\it Scientific Elite: Nobel Laureates in the United States.} Free Press, 1977.


\end{thebibliography}
\end{document}